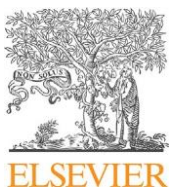
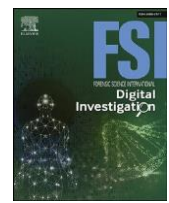

# A Comprehensive Analysis of the Role of Artificial Intelligence and Machine Learning in Modern Digital Forensics and Incident Response

Dipo Dunsin [a], Mohamed C. Ghanem [a & b], Karim Ouazzane [a], Vassil Vassilev [a]

[a] *Cyber Security Research Centre, London Metropolitan University, London, N7 8DB, UK*
[b] *Department of Computer Sciences, University of Liverpool, Liverpool L69 3BX, UK*

**Abstract**

In the dynamic landscape of digital forensics, the integration of Artificial Intelligence (AI) and Machine Learning (ML) stands as a transformative technology, poised to amplify the efficiency and precision of digital forensics investigations. However, the use of ML and AI in digital forensics is still in its nascent stages. As a result, this paper gives a thorough and in-depth analysis that goes beyond a simple survey and review. The goal is to look closely at how AI and ML techniques are used in digital forensics and incident response. This research explores cutting-edge research initiatives that cross domains such as data collection and recovery, the intricate reconstruction of cybercrime timelines, robust big data analysis, pattern recognition, safeguarding the chain of custody, and orchestrating responsive strategies to hacking incidents. This endeavour digs far beneath the surface to unearth the intricate ways AI-driven methodologies are shaping these crucial facets of digital forensics practice. While the promise of AI in digital forensics is evident, the challenges arising from increasing database sizes and evolving criminal tactics necessitate ongoing collaborative research and refinement within the digital forensics profession. This study examines the contributions, limitations, and gaps in the existing research, shedding light on the potential and limitations of AI and ML techniques. By exploring these different research areas, we highlight the critical need for strategic planning, continual research, and development to unlock AI's full potential in digital forensics and incident response. Ultimately, this paper underscores the significance of AI and ML integration in digital forensics, offering insights into their benefits, drawbacks, and broader implications for tackling modern cyber threats.

## 1. Introduction

In recent years, the field of digital forensics has expanded rapidly, relying on technology to collect and analyse digital evidence during criminal investigations, in accordance with Casey (2011). As the use of digital evidence in criminal investigations continues to rise, there is a greater need for efficient and effective crime investigation strategies. Machine learning (ML) and artificial intelligence (AI) are two potent technologies that have the potential to revolutionise digital forensics by enabling analysts to process vast amounts of data swiftly and precisely, thereby detecting crucial evidence, as stated by Du et al., (2020).

This research paper will begin by providing an overview of the field of digital forensics and the challenges that digital forensic analysts face, including the sheer volume of data, the variety of digital devices, and the dynamic nature of the digital world. The paper will then examine the current use of AI and ML in digital forensics and the obstacles it encounters, such as the lack of standardisation and interpretability issues. Also, this paper will explore several ways in which AI and ML can be utilised to improve the efficiency and accuracy of digital forensic analysis based on image and text analysis, network analysis, and machine-assisted decision-making. Lastly, the challenges and limitations of using AI and ML in digital forensics will be discussed, as well as potential future research directions, discussions, and findings.

The use of digital forensics in criminal investigations has emerged as a burgeoning area of interest. This new field requires intensive computing to acquire, process, and analyse enormous quantities of data, making the process laborious and time-consuming. To address this challenge, Dunsin., et al. (2022) propose a variety of applications and the implementation of artificial intelligence (AI), such as how AI techniques can be applied in the field of disaster response (DF) and in the context of incident response in a constrained environment. Notably, the use of AI in criminal investigations is essential, especially given the increasing prevalence of technology and cybercrime. Numerous studies have shown that electronic-based cybercrimes constitute the vast majority of offences, highlighting the significance of a digital solution as outlined by Qadir and Noor (2021). Even though databases for storing solved, unsolved, and pending cases are growing in size, it is necessary to maintain this information online for the sake of accessibility and security. For this reason, it is natural to utilise AI and machine learning (ML) applications to train datasets that digital forensics investigators can broadly utilise.

According to Thagard (1990), human experts currently conduct forensics investigations using a variety of tools and script-based applications, which call for time and expertise and are prone to human error. In light of this, the introduction of AI technology has the potential to resolve these obstacles and enhance the investigative process's efficiency. With AI, digital forensics will be quicker, more accurate, and more streamlined, as algorithms will be able to swiftly scan through vast amounts of data, including previously closed cases. This frees up detectives' time so they can focus on other pressing matters. Also, because AI is automated, it can create networks that give criminal investigators access to added resources regardless of location.

In today's society, the use of artificial intelligence (AI) and machine learning (ML) in criminal investigations is becoming increasingly crucial. According to a report from the Identity Theft Resource Center's 2021 Annual Data Breach Report Sets New Record for Number of Compromises - ITRC, (2022), cyberattacks have increased by 68% compared to the previous year. In digital forensics, which entails the acquisition, processing, and analysis of vast quantities of digital data for criminal investigations, AI and ML have proven particularly useful. According to Garfinkel (2010), the acquisition phase of digital forensics' lifecycle employs AI algorithms to analyse complex data sets that would be impossible for a human forensic expert to do manually. As a result, AI and ML have aided the criminal justice system in solving complex digital forensic investigation issues. Cabitza, Campagner, and Basile (2023) state that despite the significant progress made with AI and ML, there are still numerous challenges in this field, such as the incompatibility of existing applications. Moreover, the acquisition and reconstruction of data for the identification of criminal acts may violate privacy laws, posing a moral and legal challenge in this field. In order for digital forensics to keep up with perpetrators, it is crucial to develop more agile and effective tools capable of overcoming these obstacles.

The deployment of artificial intelligence (AI) in digital forensics requires careful consideration of the validity of the data being analysed and processed. However, determining the validity of the data presents a significant challenge because few researchers have shared effective methods for validating the data. According to Quick and Choo (2014), assessing the value of processed data to enable researchers to reduce, compress, or duplicate large datasets during investigation and analysis is a challenge associated with the use of AI in digital forensics. It can be difficult to convey the value of data due to the fact that different cultures use different communication styles. Moreover, given their cultural backgrounds, various communities may place varying degrees of importance on a variety of factors. Moreover, according to Mohammed et al. (2019), forensic studies of digital data have not been sufficiently diversified, and the majority of cybercrime investigators have concentrated on cases involving popular Western culture. Nevertheless, dedicated machine learning of data from various regions and cultures could improve AI's ability to work with diverse groups and datasets. As well, a more diverse group of researchers could play an important role in resolving these issues.

### 1.1. Research Aims

This research is dedicated to exploring the integration of AI and ML techniques within the context of digital forensics and incident response. It places a strong emphasis on the potential these techniques hold for automating various phases of digital forensics investigations. Concurrently, the research confronts challenges associated with data volume and device diversity, assesses the practical applications of AI and ML, and investigates potential trajectories for future research. Importantly, the research takes on the substantial challenges that arise from applying AI and ML in the field of digital forensics. These challenges encompass critical concerns related to data privacy, security, data quality, and data integrity. They include things like managing data well to stop privacy breaches and intrusions, dealing with the effects of biased data that can lead to unfair or wrong results, and figuring out how to train and test ML models in the field of digital forensics when data is incomplete, disorderly, or biassed.

To conduct this study, the research will rely on a meticulous literature review, drawing extensively from scholarly and peer-reviewed sources. This approach aims to develop a profound understanding of the subject matter, assess the various factors influencing the research problem, and propose well-informed solutions. Also, the research will explore the broader implications of its findings and identify promising avenues for future research, including the exploration of emerging technologies and evolving methodologies.

### 1.2. Research Contributions

The research paper provides a valuable set of contributions to the field of digital forensics and AI/ML integration. These contributions encompass various dimensions of this field, furnishing both theoretical and practical insights that can serve as guiding lights for future research and practical endeavours.

First and foremost, the paper delivers a comprehensive overview of the contemporary landscape of AI and ML in digital forensics. It effectively addresses the formidable challenges that confront forensic analysts operating within the swiftly evolving digital milieu. These challenges encompass the formidable task of managing vast data volumes, contending with a multitude of digital device types, and grappling with the dynamic nature of digital interactions. The paper adeptly unravels the intricacies and multifaceted nature of these challenges, establishing a solid foundation for a nuanced comprehension of the roles played by AI and ML in this domain.

A substantial contribution to this research lies in its meticulous analysis of the practical application of AI and ML techniques across various facets of digital forensics. The paper ventures into the domain of how these technologies can significantly enhance the practice of digital forensic analysis. It showcases their potential for boosting efficiency and accuracy, particularly in domains such as image and text analysis, network analysis, and machine-assisted decision-making. Importantly, this exploration isn't purely theoretical; it is firmly rooted in real-world instances and case studies, providing tangible evidence of the benefits accruing from the integration of AI and ML into digital forensic practices.

Furthermore, the paper confronts the formidable challenges and limitations entailed in the utilisation of AI and ML in digital forensics. It casts a spotlight on critical issues, including concerns related to data privacy, security, data quality, and data integrity. These concerns assume paramount importance in the context of forensic investigations, with far-reaching implications for the credibility and admissibility of digital forensic evidence in legal proceedings. The research methodology adopted, a comprehensive literature review, stands as yet another substantial contribution. By drawing extensively from scholarly and peer-reviewed secondary sources, the paper achieves the synthesis of a wide spectrum of perspectives and findings. This approach not only accentuates the current state of research but also identifies gaps and outlines future trajectories, thus enriching the ongoing discourse within the field.

### 1.3. Research Motivation

There has been significant interest in the use of artificial intelligence (AI) and machine learning (ML) in digital forensics in recent years. This is due to the fact that current human expertise procedures are time-consuming, error-prone, and incapable of handling the vast quantities of forensic data that modern digital devices generate. According to Stoney and Stoney (2015), one compelling reason for this integration is to enhance the efficacy and overall performance of forensic examination. On top of that, Jarrett and Choo (2021) state that by automating and streamlining various actions involved in reviewing digital evidence, such as data analysis, image and video processing, and pattern recognition, digital forensic investigators can swiftly analyse massive amounts of data, identify pertinent information, and establish connections that may not be discernible using conventional human methods.

Guarino (2013) noted that the incorporation of AI and ML in digital forensics has grown in significance due to their potential to improve investigational precision and consistency. Moreover, Ngejane et al. (2021) reported that by training digital forensic tools to recognise specific patterns or characteristics that indicate certain types of behaviour, ML algorithms can reduce the number of false positives and improve overall accuracy. Moreover, some AI and ML algorithms, for instance, can detect patterns and anomalies that may not be immediately apparent to the human eye, which can be particularly advantageous when identifying concealed or disguised evidence, resulting in more accurate and reliable results.

AI and ML, as previously stated by Hemdan and Manjaiah (2017), can aid in digital forensic analysis by identifying anomalies in network traffic, detecting malware, classifying files based on their content, and recognising objects and people in images and videos. Another crucial application of AI and ML in digital forensics is their ability to enhance investigation consistency and identify new criminal trends. James et al., (2021) stated that machine learning (ML) models can be trained on data sets and use statistical learning to predict new data sets, allowing for the identification of new evidence and cases to investigate and reducing the number of cases that must be manually analysed.

### 1.4. Research Context and Scope

The purpose of the present study is to assess the current state of research on the application of artificial intelligence (AI) and machine learning (ML) to digital forensics and incident response tasks. Specifically, the investigation will examine the techniques and methods used to employ AI and ML for a variety of tasks, such as data analysis and triage, incident detection and response, forensic investigation and analysis, network security, and cyber security. The comparative analysis will also consider the advantages and disadvantages of deploying AI and ML in various contexts, such as bias, precision, and interpretability. The analysis will incorporate a thorough evaluation of the legal and ethical implications of employing AI and ML in digital forensics and incident response.

### 1.5. Research Challenges

The application of AI and ML in digital forensics presents a number of significant research challenges that demand scholarly attention. In 2018, Losavio et al., identified data privacy and security as one of the primary challenges that must be carefully managed to prevent intrusions and privacy violations during digital forensics investigations. The quality and integrity of data may be compromised during data collection and analysis, resulting in unreliable and inaccurate outcomes. Moreover, the presence of data bias and discrimination may result in unjust or inaccurate outcomes, highlighting the significance of ensuring unbiased training data. According to Zhang et al., (2018), the availability and quality of data can pose challenges for training and evaluating machine learning models in digital forensics, where incomplete, chaotic, or biassed data can present difficulties.

Brkan and Bonne (2020) stated another challenge involving the interpretability and explainability of ML, which can be considered "black boxes" and difficult to explain in court situations where evidence must be presented and justified.

Mohammed et al., (2016) mentioned another challenge that pertains to scalability and performance, where processing massive volumes of data generated in digital forensics investigations is a significant issue that requires the optimisation of AI and ML algorithms. Furthermore, the lack of clear standards and best practices for using AI and ML in digital forensics poses an extra challenge for digital forensics experts, as it can be challenging to determine the most appropriate techniques for a particular investigation. Relatedly, the inability to interpret and explain machine learning models poses a significant challenge for digital forensics experts, as the findings and conclusions of such models may be difficult to articulate.

Due to the diverse array of devices, operating systems, and file format types encountered in digital forensics investigations, generalisation is another significant obstacle. According to Krizhevsky et al. (2017), machine learning models may struggle to generalise across these diverse categories of data. Lipton (2018) noted that it may be difficult to determine whether the model's outputs are accurate and reliable when machine learning techniques are employed, especially in complex and ambiguous situations. As a result, as machine learning becomes increasingly important in digital forensics, it is essential to be aware of potential adversarial attacks, in which an attacker generates inputs intended to confuse machine learning models, as noted by Biggio et al., (2013). Moreover, adversarial attacks are especially worrisome in digital forensics because the stakes are high and the implications of inaccurate or unreliable results could be severe.

### 1.6. Research Approach

This research paper will use a comprehensive literature review to understand the research problem and potential solutions. It will synthesise and present information, highlighting key issues and assessing the appropriateness of solutions. The paper will also evaluate strengths and weaknesses, identify future research directions, and provide an in-depth analysis of the research strategy, findings, and recommendations. The methodology will be rigorous and systematic, providing a holistic view of the research.

### 1.7. Research Methodology for Literature Investigation

The research methodology approach for identifying the relevant literature used in this research encompasses several methodical steps to ensure the comprehensiveness and relevance of the gathered materials. This methodology involves a combination of systematic literature search, critical appraisal, and thematic categorization of the selected studies.

The first step involves defining clear objectives and the scope of the literature review. The primary objective is to explore how AI and ML are applied in digital forensics and incident response. This includes understanding the types of applications, their effectiveness, challenges faced, and future prospects. The scope is confined to scholarly articles, conference proceedings, and reputable industry reports published within the last couple of years to ensure the relevance and timeliness of the information.

The literature search is conducted across multiple London Metropolitan University academic databases, such as IEEE Xplore, ACM Digital Library, SpringerLink, and Google Scholar. Keywords and phrases like "Artificial Intelligence in Digital Forensics," "Machine Learning in Incident Response," and "AI/ML Applications in Cybersecurity Forensics" are used. Boolean operators (AND, OR, NOT) are employed to refine the search results. The use of clear inclusion and exclusion criteria is intended to filter the literature. Inclusion criteria involve factors such as the publication date, the relevance of AI/ML in digital forensics, and the academic credibility of the source. Exclusion criteria include non-English publications, redundant studies, and papers not peer-reviewed.

Data extraction involves systematically collecting information from the selected studies. This includes authors, publication year, research objectives, methodologies, findings, and key conclusions. Tools such as Mendeley or Zotero are used for reference management and to organise the literature efficiently. Each selected paper undergoes a quality assessment to evaluate the research design, methodology, data analysis, and validity of the conclusions. This step ensures that the review is based on high-quality, reliable sources. The extracted data is then subjected to thematic analysis. This involves identifying patterns and themes within the literature, such as common methodologies, findings, or gaps in research. This process helps in synthesising the data to provide a comprehensive overview of the current state of AI and ML in digital forensics and incident response.

The final step involves synthesising the findings from the thematic analysis into a cohesive narrative and comparative table. This includes discussing the prevalent trends, potential applications, challenges, and future directions of AI and ML in the field. The review aims to provide a critical evaluation of the literature, identifying areas of consensus, divergence, and unexplored territories in the research landscape. The research approach culminates in a conclusion that not only summarises the findings but also provides recommendations for future research and practical applications in the field. This includes identifying gaps in the current literature and suggesting how future studies can address these gaps. This methodology ensures a comprehensive, systematic, and unbiased review of the literature, providing valuable insights into the role of AI and ML in enhancing digital forensics and incident response capabilities.

### 2. Literature Review and Research Gaps

The digital forensic evidence life cycle is a complex and multifaceted procedure comprised of several interdependent phases. These stages include identification of data sources, collection, preservation, examination, analysis, and presentation. To acquire a comprehensive understanding of this procedure, it is imperative to meticulously investigate each phase individually, as depicted in Figure 1.

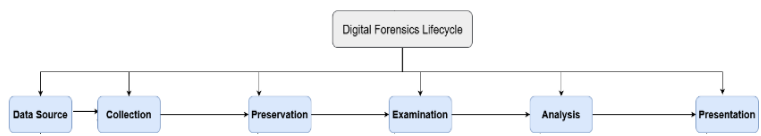

*Figure 1: Digital Forensic Evidence Life Cycle*

This paper presents a systematic literature review (SLR) as depicted in *Figure 2*, that investigates the potential of AI and ML methodologies for automating digital forensics processes. This paper's SLR provides detailed technical insights into the research gaps, limitations, and strengths of previous studies and suggests ways in which future research can resolve these gaps.

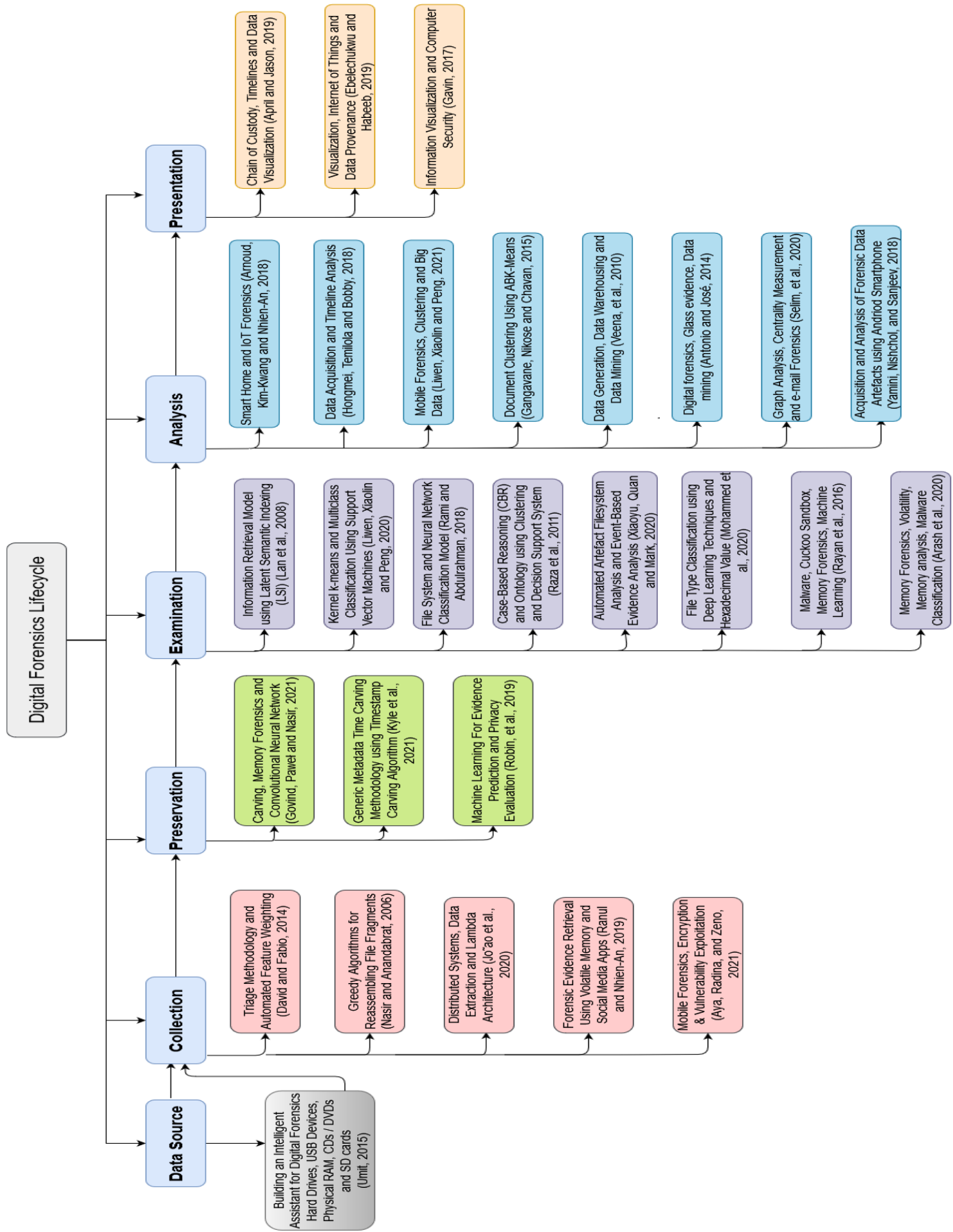

*Figure 2: Cyber Forensics Cycle-Inspired Proposed Roadmap for Systematic Literature Review (SLR) of AI and ML Techniques in DFIR*

## 2.1. Big Data Digital Forensic Investigation

According to Song and Li (2020), the widespread adoption of the Internet has led to a significant increase in cybercrime, which poses a grave threat to safety, social and economic development, and critical infrastructure. As depicted in *Figure 3*, the research presents a practical framework for conducting digital forensic investigations utilising big data technologies that manage all aspects of data collection, processing, analysis, and presentation while incorporating the most effective and cost-effective solutions. In the fight against cybercrime, the study contributes considerably to the fields of digital forensics and big data analytics.

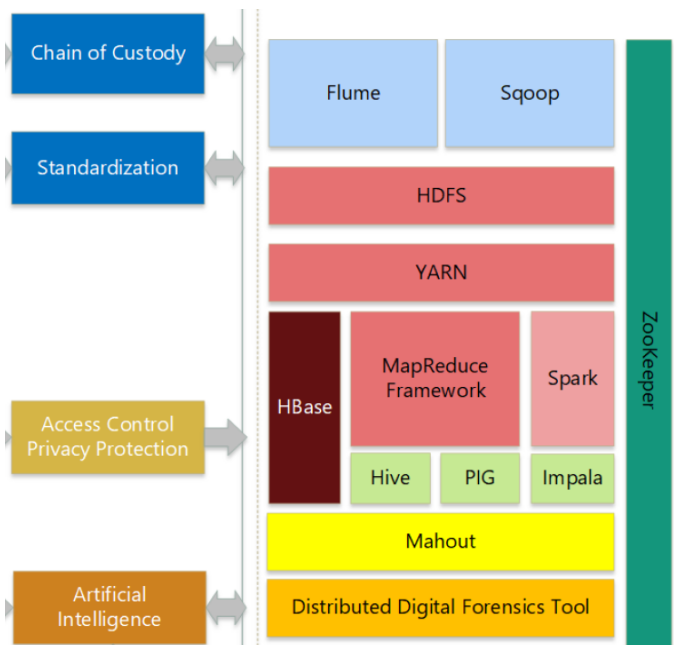

*Figure 3: Big Data Digital Forensics Framework by Song and Li's (2020)*

However, the research does not address the issue of the validity of the data being processed in the preservation process of investigations involving big data. This is a significant challenge, as people from different parts of the globe present different types of data and obstacles, and the forms may vary across devices and platforms. Besides, distinct cultural backgrounds may influence the significance and meaning of data when compared to Western languages. That being the case, it is essential to analyse and differentiate the data using the same artificial intelligence technology, which the research neglected to mention.

Despite this limitation, Song and Li's (2020) proposed framework model is robust, primarily because it takes into account the potential volume of big data and proposes advanced tools and methods for organising, standardising, and compressing the data in order to reduce the labour and cost of the process. As a result, this expedites the investigation and reduces the financial burden, given the volume of data and the rate at which it is produced. As a bonus, the framework takes into account the investigation and presentation processes, as well as how to ensure the validity, precision, security, and legitimacy of the big data investigation. Hence, this reduces the likelihood of errors, ensures dependability, and increases the likelihood that users will receive the intended results in response to their commands.

To improve the efficacy of forensic data investigations, it is crucial to avoid the inefficient use of time that frequently results from sifting through vast amounts of data without sufficient guidance or an adequate comprehension of the user's goals. Prior to initiating an investigation, it is crucial to ensure that the instructions and objectives are explicitly and exhaustively defined in order to achieve more accurate and meaningful results. Also, using artificial intelligence systems that are tailored to the region, the user's specific goals, and the values of the targeted data sources can make digital forensic investigations much more accurate and efficient than when standard intelligence systems are used alone. In this regard, Song and Li's (2020) research contributes to the advancement of digital forensics science by providing valuable insights into the use of big data technology to support cybercrime investigations, prevention, and online social interactions.

## 2.2. Volatile Memory Evidence Retrieval

Thantilage and Le Khac (2019) proposed a model for extracting memory dumps from RAM, as depicted in *Figure 4*, in order to acquire forensic evidence, with the primary goal of demonstrating that social media and instant messaging artefacts can serve as evidence for investigators. The authors also sought to elaborate on the nature of memory samples retrieved from RAM and their utility for digital forensics examiners and researchers. The authors refer to the challenge of extracting volatile memory (RAM) data that contains evidence from various social media and messaging platforms. Volatile memory is temporary and stores data as long as the device is powered. The difficulty arises because each social media or instant messaging app has unique ways of handling and storing data in RAM. This diversity makes it challenging to create a universal tool that can effectively extract pertinent data from all these platforms.

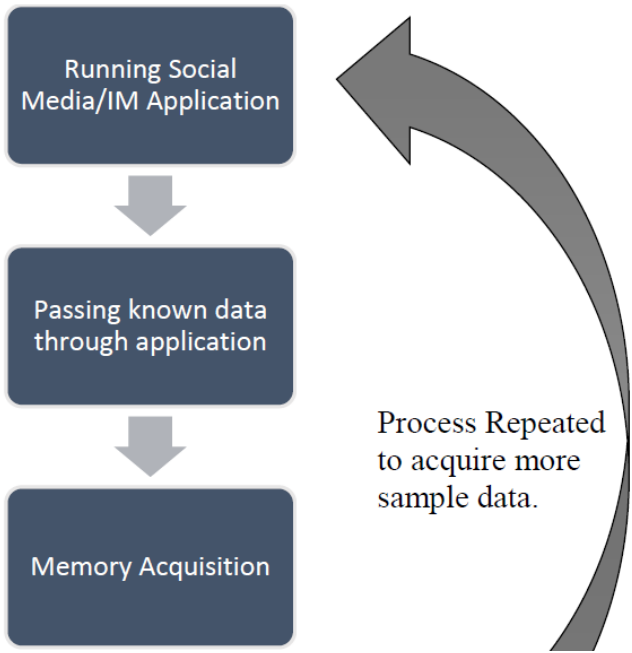

Figure 4: Proposed Framework by Thantilage and Le Khac (2019)

Indeed, the layout and organisation of data in RAM depend on the internal workings of each application, which vary not only among different applications but also across different versions of the same application. This variability presents significant challenges in creating filters or tools that can consistently and accurately interpret data from these applications when stored in volatile memory.

Thantilage and Le Khac's (2019) proposed framework structure and functions include nine phases to assure the credibility of the evidence retrieved. The authors emphasised the significance of recovering RAM data as soon as possible and avoiding restarting the computer in order to avoid losing crucial evidence. The study suggested two software programmes, DumpIt for Windows and OSXpmem for Mac OS, to retrieve memory data. DumpIt was selected due to its user-friendliness and rapid memory data acquisition, but as per the paper, DumpIt was originally limited to systems with up to 4 GiB of RAM due to its 32-bit architecture. However, newer versions of DumpIt have overcome this limitation, making it suitable for modern systems with larger RAM capacities.

OSXpmem retrieves data in RAW format, which is required for the proposed framework to produce accurate results. According to Kiley et al. (2008), using this tool requires the creation of separate profiles to guarantee volatility. Despite this, the paper fails to mention that users will be required to download a kernel that will operate concurrently with the framework in order for the extracted data to remain uncorrupted. Without the added utility, however, the memory dump may request the necessary access permissions, rendering it ineffective.

According to Yusoff et al., (2017) the framework proposed by Thantilage and Le Khac (2019) included a REGEX-based string search for the memory dump, which supports most programming languages but is not suitable for complex recursive data formats such as XML and HTML. Despite this limitation, the framework was experimentally tested on different social media and messaging platforms and operating systems, successfully retrieving valuable data that examiners could use in an investigation, including usernames and passwords for specific social media accounts. However, Thantilage and Le Khac (2019) should expand their investigation to include mobile devices and other smart home appliances.

### 2.3. File Type Identification

Mittal et al., (2021) have contributed to the field of data carving and memory forensics by presenting a new identification method for files, as depicted in Figure 5. The research aimed to demonstrate the superiority of their tool, FiFTy, compared to older file-identifying tools. The research emphasises the advantages of FiFTy, including diversified and reliable 75 file-type datasets, faster processing, higher accuracy, and better scalability. The research, however, ignored the application of data-type classification and concentrated solely on the classification of commonly used file types. Although classification of data types would have required more complex combinations, it would have been beneficial to compare FiFTy's performance to that of other data carving tools.

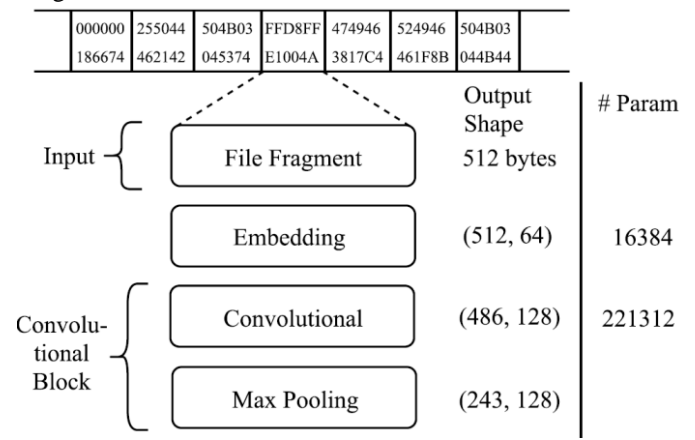

Figure 5: The Proposed Network Architecture by Mittal et al., (2021)

The 75 file-type datasets used in the Mittal et al., (2021) study had dependency issues, which made it difficult for the classifier to generalise and study images embedded in other file types such as PDF, PPT, and DOC. Then again, the study only looked at photo and graphic data from newer files that are common on SD cards used in modern IoT (Internet of Things) devices. It didn't look at a lot of different types of data in both old and new formats. In spite of this, the authors have contributed a robust research model by comparing their methodology to three other strong baselines to obtain a more objective comparison.

The research's strength resides in its exhaustive and detailed comparison of FiFTy to numerous baseline methods, as well as its extensive use of file-type datasets. This study investigated the various techniques utilised by various data carving tools for reassembling and recovering data files. It was discovered that FiFTy is a more efficient and trustworthy tool than others because it can perform multiple functions that were previously performed by multiple tools. However, the study could have specified the effectiveness of the file-type identification methods used on fragmented versus non-fragmented file structures. According to Sari and Mohamad (2020), file carving tools operate differently on fragmented and non-fragmented file structures, and only a limited number of tools are capable of recovering fragmented files.

According to Carrier (2005), *Foremost* and *PhotoRec* are a sequential file-carving tool that operates on a non-fragmented file and can identify the start and end points of files based on known file signatures, headers, and footers. However, PhotoRec can effectively recover various file types (like JPG, PNG, and DOC) by recognising their standard headers and footers. On the other hand, *Scalpel* is an advanced carving tool with fragment handling that is designed to handle fragmented files more effectively. It uses sophisticated algorithms to identify and reconstruct file fragments, and it can be configured to search for specific file types and use complex rules to reassemble fragmented files.

Most importantly, Mittal et al., (2021) research provides invaluable insights into the creation of a new instrument, FiFTy, for file identification in data carving and memory forensics. Even though data-type classification wasn't used, the large file-type datasets used and the thorough comparison of FiFTy to many baseline methods are important contributions to the field. In contrast, the limitations of the methodology include the dependency issues of the 75 file-type dataset and the emphasis on modern and current files in the selection of photographic and graphic data. According to Teimouri et al. (2020), comparing FiFTy to other robust baselines provides an unbiased assessment of its efficacy and dependability.

**2.4. Neural Network-Based Classification**

As illustrated in *Figure 6*, Mohammad's (2018) research focuses on the use of neural networks to analyse and derive conclusions from retrieved data for digital forensics in criminal investigations. The research has contributed to the reconstruction of the events leading up to the crime under investigation and the retrieval of crucial information from data such as cookies, log files, and web browser history. However, one of the limitations of his method is that data must first be transformed by third-party applications, which can be expensive and not scalable for large data volumes. Alternatively, the paper suggested that Machine Learning can address this issue by explicitly analysing data sets..

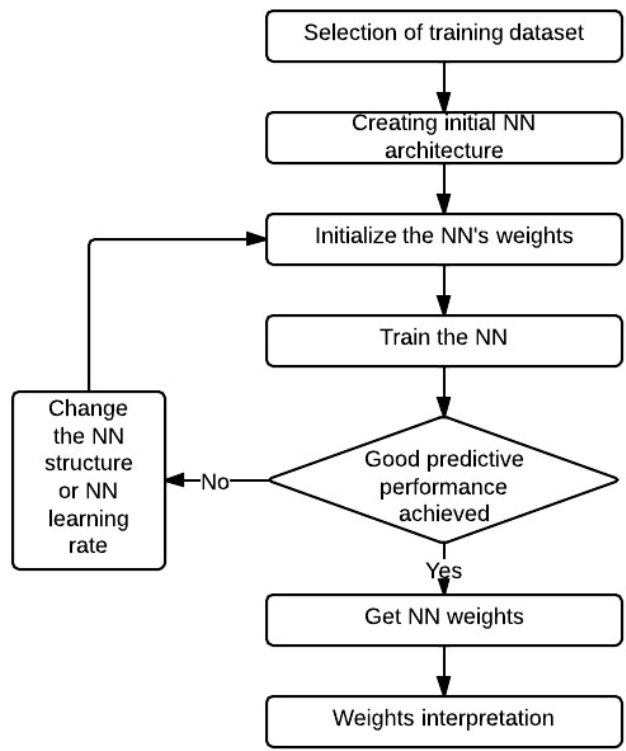

*Figure 6: Digital Forensic Classification Model by Mohammad (2018)*
.

The objective of this study is to determine if neural networks are capable of identifying and tracing the history of events to determine if other applications have modified the files. Mohammad's work expands on Palmer's (2001) nine-step framework for digital forensics and proposes a finite-state machine model with ontology to facilitate the reconstruction of historical events based on the gathered data. According to Chabot et al. (2014), one of the limitations of the proposed model is that it treats events as instantaneous occurrences rather than intermittent ones, which may cast doubt on the validity of the acquired dataThe research proposes using neural network technology to determine whether or not files have been altered and whether or not the trained datasets can accurately reconstruct past events. During this process, however, criminals may readily manipulate the models used to generate features, leading to inaccurate data retrieval. In light of this limitation, the research produced robust models using the machine learning algorithm, despite the fact that the tool used to manage small datasets may run out of memory when processing large volumes of data

The experimental results demonstrated that the created feed-forward model produced substantially satisfactory outcomes with an error rate of 10.07 percent across four distinct scenarios. However, using a single algorithm to execute multiple applications may result in system overlap and invalid results. Thus, it is of the utmost importance to develop alternative algorithms that produce accurate results. The research contributes to the advancement of digital forensics by providing valuable insights into the use of neural networks and machine learning while acknowledging the limitations and challenges that must be overcome.

## 2.5. AI-Based Incident Response

As depicted in *Figure 7*, Hasan et al., (2011) propose a computer model that uses artificial intelligence to expedite forensic investigations and reduce the time and resources needed by crime investigators. The strengths of the proposed model include its ability to efficiently analyse crime scene evidence and generate accurate conclusions. Not to mention, the use of a specialised software tool known as "chain of custody" guarantees the security of evidence and vital information, which can be stored in a database and used as a training source for the model.

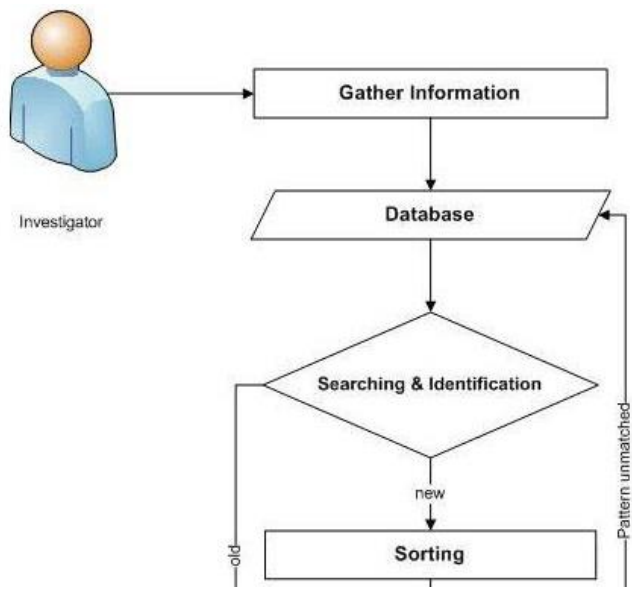

*Figure 7: Proposed Model System by Hasan et al., (2011)*

Nonetheless, the proposed model contains certain flaws. According to Nila et al., (2020), the extensive work required to collect enough data to train the model is inadequately described, and the data collected from various police agencies in the United Kingdom may be vulnerable to malware, anomalies, and malicious code injections. On the other hand, the research concisely emphasises the significance of continuously training the model to ensure accuracy, given that AI systems require substantial data inputs for training. Failure to do so can result in falsified facts and findings, leading to incorrect conclusions, according to the research.

According to Trifonov et al., (2019), Hasan et al., (2011) research method fails to account for the risk of interference from hackers and third-party software, which is a significant challenge. The model should include a method for detecting unauthorised system access, such as Intrusion Detection Systems (IDS), Intrusion Prevention Systems (IPS), Log Analysis, User Behaviour Analytics (UBA), Multi-Factor Authentication (MFA), Audit Trails and Monitoring, Port Scanning and Vulnerability Scanning, Honeypots, Anomaly Detection, and File Integrity Monitoring.

Despite the challenges, the proposed model makes significant contributions to forensic investigations. The distinctiveness of the Hasan et al., (2011) model resides in its capacity to predict any crime and to adapt and learn independently to solve new and future crimes. With sufficient resources, the proposed improvement in contemplating criminals' psychology can be a valuable accumulation for collaboration with behaviour analysts. This can establish a pattern that can be added to grouped data sets to assist with crime prevention and resolution.

## 2.6. Automated Artefact Relevancy Determination

As depicted in *Figure 8*, Du et al., (2020) investigated the possibility of utilising previously investigated digital forensic cases to aid in the investigation of new digital forensic cases using automated artificial intelligence systems. Their research was specifically intended to rank the importance of file artefacts required for forensic examination.

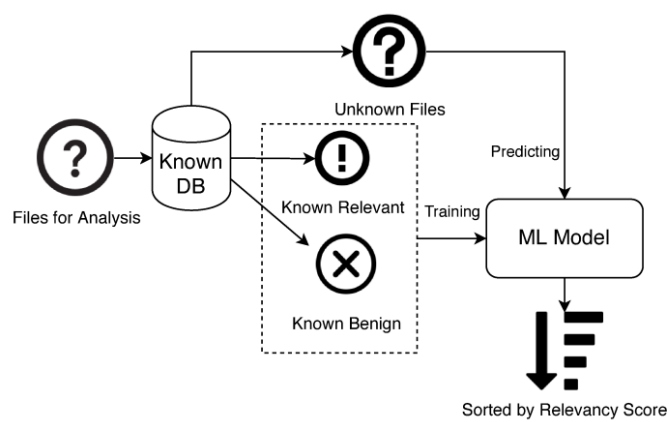

*Figure 8: Overview of the Approach by Du et al., (2020)*

By evaluating the automation process with files from three distinct case scenarios to identify similar and unknown files, the study accomplished its objective. The researchers demonstrated the advantages of using trained automated processes to examine file evidence, such as significant time savings and the avoidance of negative psychological effects that human investigators may experience when examining distressing evidence.

The researchers acknowledged the significance of timeline analysis in identifying and ordering events in chronological order. However, the research did not investigate how the applicability of known similar files and novel evidence would be weighed and prioritised in relation to new cases. Likewise, the study did not account for situations in which there are fewer known file artefacts available for machine learning training, which may impact the approach's efficacy.

Although the research approach had some limitations, such as the difficulty in identifying "interesting" files and the possibility of overfitting, the study's model was robust because it was validated using experimental data from multiple case scenarios. The use of multiple scenarios by the researchers is an excellent method for preventing bias in future research. To avoid overfitting, future research should avoid using machine learning models with very few features. Moreover, Du et al., (2020) research could utilise disc images from actual past cases with permission from officials involved or generate experimental data with information similar to real cases to obtain better research outcomes as opposed to fabricated experimental data, which may lead to more false negatives and false positives, as stated by Eykholt et al., (2018).

**2.7. Large-Scale E-mail Dataset Analysis**

As shown in *Figure 9*, Ozcan et al., (2020) emphasised the importance of email analysis as a primary source of evidence when acquiring forensically pertinent data from disc images. The objective of their research is to develop an end-to-end distributed graph analysis framework for large-scale digital forensic datasets, along with an evaluation of the accuracy of the centrality algorithms and the scalability of the proposed framework in terms of running time performance. The research proposes an algorithm-based framework that can perform the task of analysing email files more efficiently and effectively in response to the challenges posed by traditional methods for managing large volumes of email files.

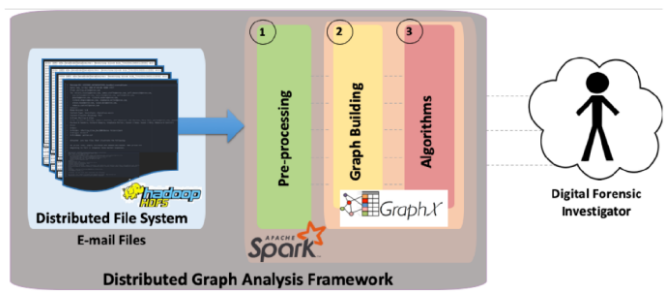

*Figure 9: Framework with innovative technology by Ozcan et al., (2020)*

The research by Ozcan et al., (2020) is robust and exhaustive, employing a controlled and empirical methodology that is critical and verifiable. The researchers developed an edge-transmitted graph methodological approach for coping with large forensic datasets, implemented it with widely adopted open-source technologies, and analysed the algorithmic precision of its nodes. The research paper presented three implementations to demonstrate the efficacy of the proposed framework, as well as experiments on an email dataset to demonstrate its superiority to conventional methods.

One limitation of Ozcan et al., (2020) research methodology is the framework treatment of email addresses in the dataset as originating from distinct individuals, which fails to account for the possibility that multiple email addresses belong to the same individual. To increase the accuracy with which prospective offenders are identified during forensic investigations, the pre-processing phase should be modified to permit the matching of email addresses. Notably, the research employs a secure and efficient local testing environment with high-performance computing resources, which increases the tests' credibility.

A future study could be enhanced by utilising multiple email datasets to evaluate the framework and avoid biassed results. While Ozcan et al., (2020) utilised the Enron email dataset, which is one of the largest and most exhaustive collections of meaningful emails, it may be incomplete due to the fact that it only contains messages from users who were employees of the same company. Balayn et al., (2021) discourage the use of unique, distinct, and trained datasets in testing experiments because they are likely to be unjust to groups that are not included in the dataset. On the other hand, the use of diverse email datasets would introduce a variety of cases from diverse groups, cultures, and situations that would further demonstrate the framework's dependability.

Lastly, Ozcan et al., (2020) research makes a significant contribution to the field of digital forensics by introducing an effective framework for evaluating the accuracy and scalability of large forensic datasets. Although their methodology has limitations, their diligence and empirical approach guarantee the study's reliability and validity.

**2.8. Data Mining Methods**

In their study, Tallón-Ballesteros and Riquelme (2014) examined various data mining techniques applicable to digital forensic tasks, focusing on glass identification as an illustration of a data problem. Digital data analysis is a quicker and more accurate method for evaluating large volumes of data than traditional forensic analysis through lab experiments, which can be challenging and expensive. To accomplish their research objective, the researchers employed the "stratified four-fold cross-validation" method, which involved dividing the existing dataset into four equal parts and analysing individual training sets.

The study acknowledged that statistical analysis can be used to identify statistically significant differences in the outcomes of stochastic methods. However, non-academic algorithms with a single output cannot be subjected to statistical analysis. To study many different types of classifiers and machine learning methods in 2014, Tallón-Ballesteros and Riquelme used a strong research model. They looked at decision trees, Bayes classifiers, artificial neural networks, and rule-based classifiers. This comprehensive strategy yielded more reliable and comprehensive results. However, the research was restricted to only two analysis tests, namely Cohen's Kappa and accuracy measures, in order to evaluate the models developed by the various classifiers. Incorporating supplementary types of analysis experiments could have provided further information for comparison in the machine learning task and identified any new problems with the model or data.

As shown in *Table 1*, Tallón-Ballesteros and Riquelme (2014) obtained comparable results to Silva and Hruschka (2013) using a ten-fold cross-validation procedure on the same dataset. The ten-fold cross-validation procedure, however, lacked a statistical analysis for the stated problem. The research conducted by Tallón-Ballesteros and Riquelme (2014) highlighted the significance of having diverse results for comparison using various parameters. The results of the experiment improved after the parameters were fine-tuned, resulting in algorithmic performance that exceeded the values of the analysis measures of other experimental results with default parameters.

| Algorithm type | Classifier approach | Method | Accuracy (%) | | Cohen's kappa | |
|---|---|---|---|---|---|---|
| | | | Training | Test | Training | Test |
| Non-stochastic | Decision Tree | C4.5 | 90.50±1.59 | 68.00±8.33 | 0.8700±0.0214 | 0.5663±0.1005 |
| | Bayes | BayesNet | * | 69.59±7.50 | * | 0.5830±0.0974 |
| | Nearest neighbour | Euclidean 1-NN | 100.00±0.00 | 69.64±7.84 | 1.0000±0.0000 | 0.5867±0.1062 |
| | Nearest neighbour | Manhattan 1-NN | 100.00±0.00 | **70.13±6.85** | 1.0000±0.0000 | **0.5949±0.0973** |
| | Nearest neighbour | Chebyshev 1-NN | 100.00±0.00 | 65.04±6.18 | 1.0000±0.0000 | 0.5222±0.0849 |

*Table 1: Accuracy and Cohen's Kappa Measures for 6-class Training and Test Results by Tallón-Ballesteros and Riquelme, (2014)*

In light of this, future research evaluating data mining approaches should not only concentrate on accuracy but also consider other crucial factors such as dependability and utility. These actions would provide information regarding the experiment and reduce data mining errors.

**2.9. Metadata Analysis Using Machine Learning**

Toraskar et al., (2019) recommended in their study the use of information acquired from storage devices to analyse and detect alterations, as illustrated by *Figure 10's* output results. The study explored the potential for unsupervised machine learning classification to aid in forensic analysis. This study adds to the body of research that supports the right use of machine learning and forensic technologies for data analysis, such as SOM viability, in criminal investigations.

This study emphasises the benefits of the Encase Imager tool, the Encase Forensic Tool, and the FTK (Forensic Toolkit), which are known for their speed and efficiency in producing digital reports in CSV format. However, it's important to note that while Autopsy is a robust forensic tool, it is not particularly known for its speed. Also, the CSV (Comma-Separated Values) reports are indeed exports of data, typically tables from databases, in a raw CSV format. These reports are known for their simplicity and ease of use, as they can be opened with various software, including text editors and spreadsheet programmes.

However, the research also acknowledges the limitations of these tools, such as their inability to process non-English languages and their inability to distinguish between false negatives and genuine positives. Beyond that, FTK lacks a user-friendly interface and effective search capabilities. Nonetheless, according to Wehrens (2009), research should present an alternative method for carrying out parallel inquiries and comparing results.

The research of Toraskar et al., (2019) introduces a self-organising map (SOM) as a clustering tool for MATLAB in order to resolve these limitations. The paper presents four cybercrime scenarios to evaluate the dependability of machine learning in forensic analysis. These scenarios include data theft, corporate fraud, hacking, and document fraud. In each case, SOM is used to cluster notable artefacts, demonstrating its effectiveness in different contexts. In the study, the inputs to the SOM model were primarily based on metadata categories from the forensic cases. These categories included EXIF Metadata, Extension Mismatch Detected, Web History, and USB Device Attached Data. The SOM utilised this data to cluster and identify notable files in the four different cybercrime scenarios.

SOM is advantageous because it can readily cluster data, identify common characteristics, and handle various problem classifications. The paper suggests using various cluster sizes to guarantee accurate results.

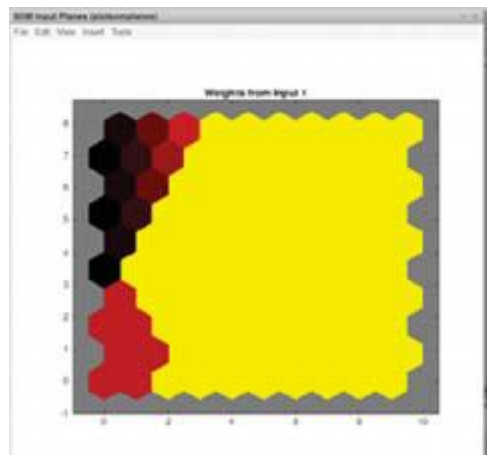

*Figure 10: SOM Outputs by Toraskar et al., (2019)*

According to Toraskar et al., (2019), the SOM mappings generated by MATLAB were clustered, demonstrating the viability of using SOM with enumerated artefacts and metadata in criminal investigations. However, the research cautions that it may be difficult to acquire a perfect mapping if the groupings are unique. Therefore, anomalies may form, resulting in the appearance of two identical clusters in peculiar regions of the map. In spite of this limitation, the research findings are trustworthy, as the selected metadata and cluster sizes lead to accurate results.

The research of Toraskar et al., (2019) contributes significantly to the application of machine learning and forensic tools in digital forensic analysis. This study acknowledges the limitations of existing tools and introduces a novel method for clustering data using SOM. However, the research should have taken into account the difficulty of obtaining flawless mappings when groupings are distinct. Nevertheless, the findings are trustworthy and provide a firm foundation for future studies.

**2.10. Chain of Custody**

The research conducted by Tanner and Bruno (2019) proposes a valuable tool for visualising and organising data related to the chain of custody process in criminal investigations, as depicted in *Figure 11*. The objective of the research was to develop an instrument that satisfied the three fundamentals of timeline representation: literal, linear, and global timelines. The proposed implementation of the tool included HTML input and output in the form of tables and timelines, enabling examiners to efficiently manage criminal evidence.

Walny et al., (2020) state that one of Toraskar et al., (2019) research's strengths is the use of *taffy.js* as a database library, which improves the tool's performance and reduces downtime, resulting in a system that runs smoothly. On top of that, the offline feature of the application protects the system from online malware and hackers, while the *.csv* export feature enables the secure storage of data. Also, the tool's user-friendly characteristics make it simple to use and interactive.

However, Tanner and Bruno (2019) research has neglected some deficiencies that need to be addressed, such as the tool's visual appeal. While the *vi.js library* used to construct the timeline network has built-in behaviours, it may need to be modified to enhance the visual appeal of the user interface. On the other hand, the system's load time may be sluggish due to the use of nodes instead of clusters, resulting in an annoying *"loading... "* message for users.

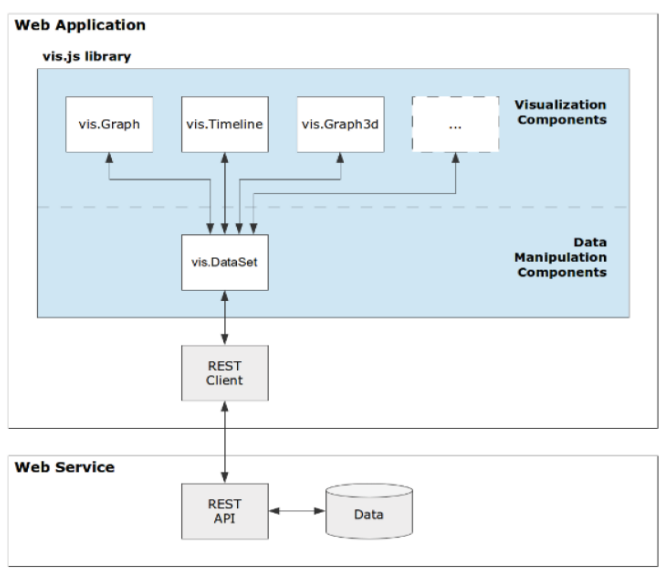

*Figure 11: Vis.js library components by (Tanner and Bruno, 2019)*

One of the major contributions of Tanner and Bruno's (2019) research is the creation of a tool that satisfies all three fundamentals of timeline representation, making it superior to existing models. The implementation of the tool would assist minor departments in eliminating the manual chain of custody process and storing information securely for an extended period of time without risk of modification. According to Elgohary et al., (2022), the paper acknowledges the need for supplementary enhancements, such as adding a search engine and a data grouping feature, to facilitate access to information and patterns in similar cases. Overall, the research was successful in accomplishing its goal of developing a chain of custody data visualisation tool based on time.

## 2.11. Memory Forensics Using Machine Learning

Through the extraction of memory images, Mosli et al., (2016) sought to develop a model for automating the detection of malware. Specifically, the study concentrated on three key malware artefacts: imported DLLs (Dynamic Link Library), malware-modified registry keys, and API (Application Programming Interface) functions, with the intention of developing a highly accurate and user-friendly model. Through experimentation, the researchers were able to accomplish their goal, with the model achieving accuracy rates ranging from 85.7% to 96%. The study emphasises the significance of using memory images in malware detection because they permit the extraction and analysis of multiple artefacts, resulting in more precise conclusions.

While there are numerous malware detection techniques on the market, Mosli et al., (2016) contend that the proposed model is preferable due to its resistance to manipulation. The extracted information is uncommon, precise, and diverse and is capable of handling millions of global malware variants. According to Sihwail et al., (2019), it is important to note that the design of the proposed model only enables the detection of already-present malware and does not prevent malware from infiltrating the system.

Mosli et al., (2016) research employed the finest feature-extraction techniques, resulting in a flawless data acquisition and feature extraction process, regardless of the volume of data being analysed, in order to develop a method that can address potential vulnerabilities in malware design. Scikit Learn is recommended as a tool for feature extraction due to its precision and user-friendliness, but it is not appropriate for data visualisation or string processing.

| Classifiers | Registry | DLLs | APIs |
|---|---|---|---|
| SVM | 94.4 | 88.7 | 92.3 |
| SGD | 96 | 87.8 | 93 |
| Random Forest | 94.9 | 90.5 | 91.5 |
| Decision Tree | 94.9 | 88.7 | 90.7 |
| KNN | 93.9 | 89 | 90.7 |
| BernoulliNB | 93.4 | 89.6 | 89.2 |
| MultinomialNB | 92.9 | 85.7 | 89.7 |

*Table 2: Summary of Accuracy by (Mosli et al., 2016)*

The research of Mosli et al., (2016) utilised seven training models to establish accuracy and generate linear, simple-to-analyse results. Regardless, using both accuracy and AUROC produced significant and conclusive results, as shown in *Table 2*. This shows that the proposed model can find malware even when there is a huge amount of data if the right tools are used. The study accomplished its objective and demonstrated that it is possible to detect malware using machine learning with the proper tools and techniques. The study suggests that future research should concentrate on identifying further memory artefacts for analysis, broadening the spectrum of data, and developing methods to detect malware before it enters the system. This can be accomplished by collecting and analysing a sufficient quantity of malware data.

## 2.12. Malware Classification using Feature Engineering

As depicted in *Figure 12*, Lashkari et al., (2021) presented VolMemLyzer, a digital instrument that enables memory analysis of live malware infections to extract feature sets for characterising malware. The research contributes significantly to the field of digital forensics by addressing multiple tasks, including malware extraction, memory dump analysis, feature extraction, feature ranking, and machine learning classification of both benign and malicious samples.

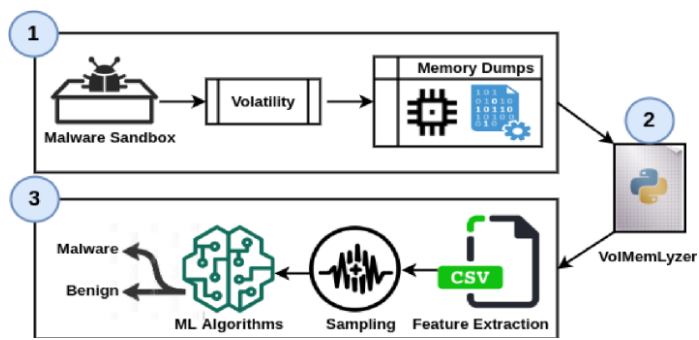

*Figure 12: Proposed Model by Lashkari et al., (2021)*

One of the research's strengths is its emphasis on the importance of memory analysis tools in identifying the specific areas affected or compromised by malware, thereby guiding digital forensics analysts as to where to concentrate their examinations. Further to this, the study acknowledges the limitations of classifiers used for analysis and classification, which can memorise training samples and produce incorrect results. To circumvent this restriction, the researchers added 7% noise to the sampled data, and by increasing the memory dump size by 100%, they were able to acquire 1900 samples in the dataset through Weka.

The research further emphasises the significance of employing multiple classifiers to achieve a more precise classification of malware families. Various classifiers, such as random forests, k-nearest neighbour, decision trees, and Adaboost, were used to identify benign samples during binary classification. According to San, Thwin, and Htun, (2019), it was discovered that Random Forest and k-nearest neighbour classifiers were more effective at classifying malware families. This is an important contribution to the discipline because it emphasises the significance of using the most effective classifiers for accurate classification.

Nonetheless, Lashkari et al., (2021) study has some limitations. The researchers did not account for the prospect of receiving fewer memory dumps for analysis after malware execution, necessitating a 100% increase in samples to obtain 1900 samples in the Weka dataset. Similarly, the research did not address the difficulty posed by contemporary malware, which employs techniques such as process hollowing to avoid detection and analysis. As a result, future research can explore methods to detect and analyse modern malware that employs process hollowing techniques and address these limitations.

Last but not least, the research conducted by Lachkari et al., (2021) is a significant contribution to the field of malware analysis, providing insight into the significance of memory analysis tools, the limitations of classifiers used for analysis and classification, and the necessity of using multiple classifiers to achieve more accurate classification.

### 2.13. Android Device Forensic Handling Insights

The paper *"An Overview on Handling Anti-Forensic Issues in Android Devices Using Forensic Automator Tool"* by Bhushan and Florance (2022) critically addresses digital forensic challenges, especially in Android devices, through a forensic automator tool. This review assesses its structure, approach, contributions, and limitations. Structured with a concise abstract, a comprehensive introduction to mobile forensics, and a focus on anti-forensic techniques, the paper effectively establishes its primary objective. It delves deeply into these challenges, culminating in a discussion on the proposed solution and future research.

The methodology showcases a thorough analysis of anti-forensic methods, highlighting mobile forensics evolution and the sophistication of these techniques. The innovative use of a machine learning-based forensic automator tool, particularly the Support Vector Machine (SVM), as depicted in *Figure 13*, used for file encoding detection, underscores the paper's technical depth.

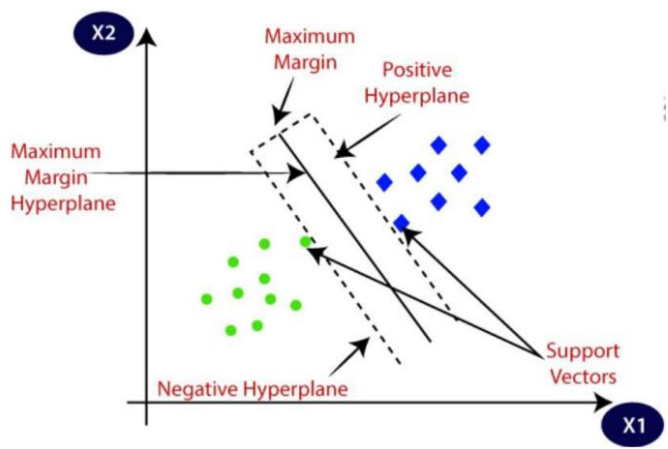

*Figure 13: The SVM Hyperplane by Bhushan and Florance (2022)*

Significantly, the paper identifies various anti-forensic techniques and their impacts, proposing a machine learning application to automate file encoding detection and decoding, which is a major advancement in forensic investigations.

Practically, the tool offers considerable implications for forensic investigators, streamlining the detection and decoding process on Android devices. The paper also provides an in-depth understanding of anti-forensic methods, which is crucial for developing stronger forensic tools. However, the paper could explore the tool's limitations and real-world application challenges more deeply, like the machine learning model's accuracy and reliability. Inclusion of case studies or practical examples would enhance the tool's demonstrated effectiveness.

Finally, future research suggestions include integrating this tool with other forensic methods and expanding its capabilities to address a broader range of anti-forensic techniques, offering promising avenues for further exploration in the field.

### 2.14. Cybersecurity Tactics: Windows Artifact Analysis

The research paper titled *"Detecting Adversary using Windows Digital Artefacts,"* authored by Liew and Ikeda (2019), investigates the detection of malicious behaviours linked with Advanced Persistent Threats (APTs) through the analysis of Windows digital artefacts as illustrated in *Figure 14*. This review critically assesses the paper's methodology, results, and overall contribution to the field of cybersecurity, underpinned by relevant academic literature.

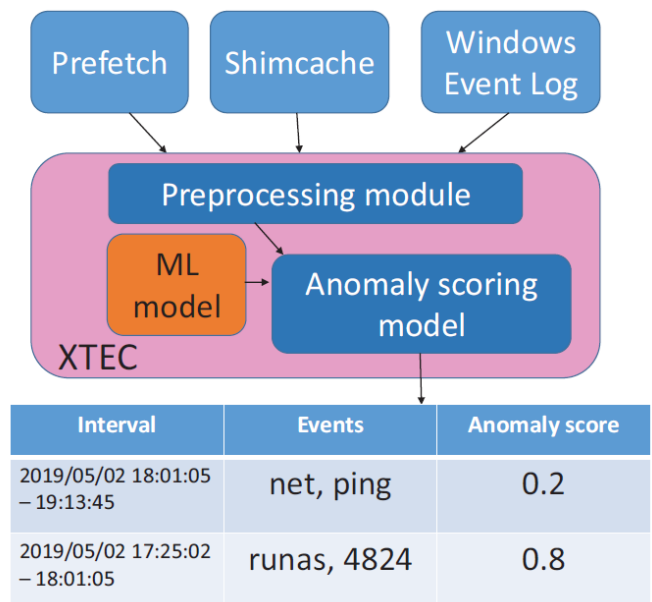

Figure 14: The architecture of XTEC by Liew and Ikeda (2019)

The authors advocate a novel technique for identifying APTs, independent of third-party sensors. Central to their approach are Windows operating system artefacts, especially the Application Compatibility Cache (Shimcache). A key innovation is their algorithm designed to estimate the execution times of files, a vital feature considering Shimcache does not inherently provide this data. This method reflects the growing trend in cybersecurity research that emphasises the importance of harnessing system-generated data for threat detection (Berlin et al., 2015; Carvey, 2018).

The paper presents an interval estimation algorithm for determining file execution times, seamlessly integrating this with machine learning strategies to forge the XTEC system. This blended approach, merging rule-based analysis with machine learning, is gaining recognition for its effectiveness within the cybersecurity sphere (Virvilis and Gritzalis, 2013). The employment of Random Forest classifiers, lauded for their robustness in varied settings, lends further weight to their methodology (Breiman, 2001).

Included in the paper is a real-world case study, serving to authenticate the efficacy of the XTEC system. Such practical testing is pivotal for demonstrating the system's real-life applicability, resonating with the need for empirical validation in cybersecurity research (Alshamrani et al., 2019). However, the case study's specifics are somewhat constrained due to confidentiality concerns, somewhat limiting the ability to thoroughly gauge the system's effectiveness across varied scenarios.

The development of a cutting-edge algorithm for estimating file execution times is the research's primary contribution, and then comes the creation of a detection system that is independent of outside surveillance tools. This advancement holds significant value in the realms of digital forensics and incident response, bridging existing gaps in these fields (Tankard, 2011; Virvilis and Gritzalis, 2013). The authors further outline prospective avenues, including enhanced data collection methods for improved model performance and advocating for the sharing of public data, underscoring a sustained commitment to propelling the field forward.

Despite the paper's pioneering approach and significant contributions to the field of cybersecurity, several areas are identified where enhancements could be beneficial, such as an expanded evaluation of the algorithm's performance and its limitations in a broader range of scenarios. This would help to confirm its applicability and effectiveness in a variety of contexts. The methodology's reliance on Windows-specific artefacts could potentially limit its effectiveness in environments where multiple operating systems are in use. Broadening its scope to include other systems could enhance its utility. Although the included case study is valuable, it falls short in providing the depth of detail needed for a more compelling validation of the system's effectiveness. Enriching the case study with more comprehensive data would strengthen the evidence for the system's capabilities.

### 2.15. Machine Learning's Role in Locked Shields Defence

The research paper titled *"Machine Learning-based Detection of C&C Channels with a Focus on the Locked Shields Cyber Defence Exercise,"* authored by Känzig et al., (2019) and Ghanem, (2022), introduces a system tailored to identify Command and Control (C&C) channels within network traffic, with particular emphasis on the Locked Shields cyber defence exercise, as seen in *Figure 15*.

The paper tackles a critical issue in cybersecurity: the detection of C&C channels, vital for the operation of botnets. Its distinctiveness stems from its application to the Locked Shields exercise, NATO's largest live-fire cyber defence exercise. This research is not only pertinent but also strives to offer a solution that is both efficient and scalable. By leveraging machine learning, the study aims to enhance the detection of malicious traffic, a key challenge in the field of cybersecurity.

The authors opt for a machine learning approach, specifically utilising a random forest classifier. This method is particularly well-suited given the nature of the data and the overarching need for efficiency and scalability in such systems. The model is trained on data derived from past cyber attacks, particularly from the 2017 and 2018 Locked Shields exercises. While this training approach is practical, there may be concerns about its ability to generalise across various network environments or against novel attack vectors that have not previously been encountered

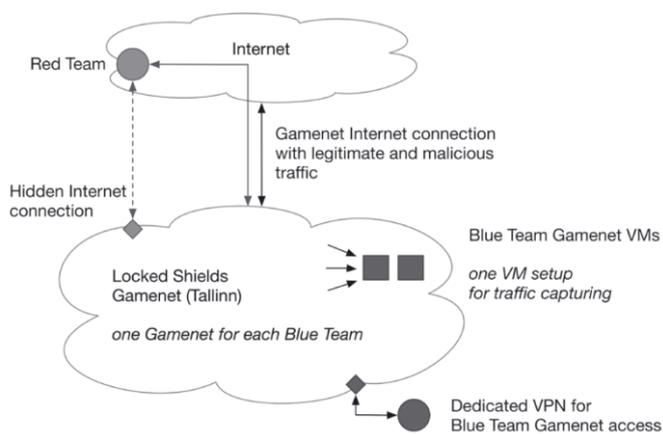

Figure 15: Locked Shields Environment Overview by by Känzig et al., (2019).

The paper excels in its comprehensive data analysis and meticulous feature selection. The authors have compiled a robust dataset, including traffic captures and Red Team logs, which provides a solid base for their machine learning model. The process of feature extraction and selection is detailed, focusing on computational efficiency. However, a deeper discussion of the reasoning behind the selection of specific features could further strengthen the paper.

In terms of results and evaluation, the system demonstrates high accuracy, claiming 99% precision and over 90% recall in near real-time. These impressive results underscore the model's effectiveness in the context of the Locked Shields exercise. However, a more thorough comparison with other existing systems would improve the paper and provide a better contextual understanding of its performance.

The practical application of this system in the context of locked shields is clearly articulated and well-supported. The authors also discuss the potential deployment of the system in similar exercises or real-world scenarios. However, a primary limitation noted is the model's potential inflexibility in adapting to different network environments or to new types of C&C communication methods not represented in the training data. It is recommended that future iterations of the dataset be expanded to include a broader range of attack scenarios and network environments. As a result, performing so will enhance the functionalities of the model. Moreover, a thorough investigation into alternative machine learning approaches, such as deep learning, could potentially unveil substantial insights concerning the complexities of C&C traffic patterns.fffff

## 2.16. Meta-Heuristic JPEG Reassembly in Forensics

The paper *"A Meta-Heuristic Method for Reassembling Bifragmented Intertwined JPEG Image Files in Digital Forensic Investigation"* by Ali et al., (2023) presents a novel advancement in digital forensics with the introduction of the Meta-Heuristic Reassemble Images (MHRI) method, as seen in *Figure 16*. This method, designed for the recovery of fragmented JPEG images, is a testament to the innovative approaches evolving in digital forensic investigations. The MHRI method is unique because it uses restart markers, the Coherence of Euclidean Distance metric (CoEDm), and a genetic algorithm with a cost function. This creates a new way to solve the difficult problem of putting back together two broken JPEG images that are intertwined.

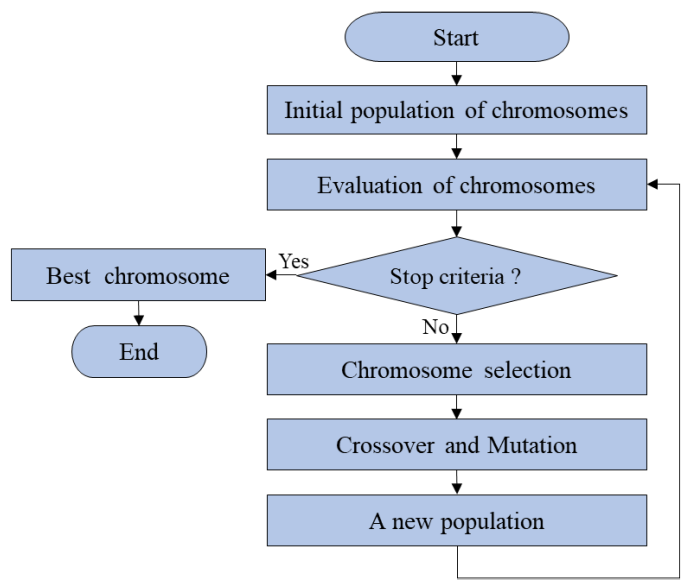

Figure 16: The flowchart of the genetic algorithm by Ali et al., (2023)

The efficacy of the MHRI method is demonstrated through extensive testing on both public and private datasets. Remarkably, it achieved a complete recovery of all bifragmented intertwined JPEG images and a 48.4% recovery rate of all JPEG images in the test sets, which is a significant improvement over existing methods such as RXmK, mK, RevIt, and XmK. This superior performance highlights the method's potential to enhance the accuracy and efficiency of digital forensic investigations. Also, the paper provides a thorough explanation of each component of the MHRI method, offering valuable insights into the complexities and intricacies involved in the process.

However, despite these strengths, the paper by Ali et al., (2023) also reveals certain weaknesses in the MHRI method. One notable concern is the computational complexity of the method. The use of genetic algorithms and multiple metrics might necessitate substantial computational resources, yet this aspect is not adequately addressed in the paper. Furthermore, the MHRI method's scope of applicability appears limited, as it is specifically tailored for bifragmented intertwined JPEG images in linear order. This specialisation may restrict its application in a broader range of forensic scenarios, particularly for different types of fragmented files or formats. The paper also lacks an analysis of the method's application in real-world forensic cases, which is crucial for evaluating its practical effectiveness and reliability. However, the usability and accessibility of the MHRI method for forensic practitioners are not discussed, raising questions about its ease of use given the inherent complexity of the approach.

Looking ahead, future research should aim to address these limitations. Efforts could be directed towards optimising the computational efficiency and speed of the MHRI method, thereby enhancing its practicality for real-time forensic investigations. Expanding the scope of the method to include a wider range of fragmented files and different file formats would significantly increase its utility in the field of digital forensics. Furthermore, applying the MHRI method to real-world forensic scenarios would provide valuable insights into its effectiveness and highlight areas for improvement in practical settings. As well as developing a more user-friendly interface and providing comprehensive training materials, this could facilitate the adoption of the method by forensic professionals.

### 2.17. Memetic Algorithms in Digital Forensics

*"Enhancing Digital Forensic Analysis Using Memetic Algorithm Feature Selection Method for Document Clustering"* by Al-Jadir et al., (2018) is a research paper that goes into great detail about how to make digital forensic analysis better by using new feature selection methods in document clustering. An illustration of the system architecture is displayed in *Figure 17*. The paper's significance lies in addressing the challenges of managing the ever-increasing volume of digital documents in criminal investigations, where efficient and accurate clustering of crime-related documents is crucial.

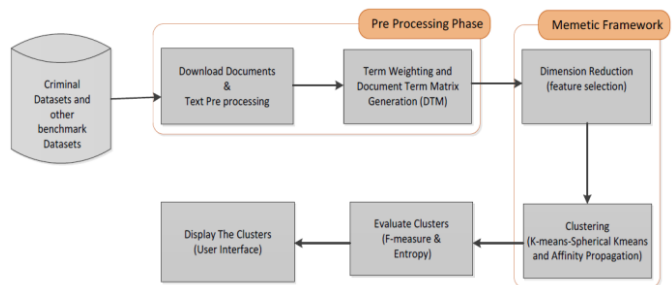

*Figure 17: System Architecture by Al-Jadir et al., (2018)*

The authors propose a Memetic Algorithm Feature Selection (MAFS) approach, combining a Genetic Algorithm-based wrapper feature selection with the Relief-F filter method. This hybrid approach is applied to enhance two clustering algorithms - k-means and Spherical k-means (Spk) - and is tested on crime reports, criminal news, and benchmark text datasets. The performance of these algorithms is evaluated based on the clustering outcomes before and after applying the MAFS method. This approach's effectiveness is demonstrated through significant improvements in the performance of both k-means and Spk algorithms after applying MAFS.

The paper's strengths include a well-structured methodology, a clear explanation of the proposed MAFS method, and thorough experimental results. The use of both crime-related and benchmark datasets ensures robust validation of the proposed method, showing its applicability and efficiency in real-world scenarios (Al-Jadir et al., 2018). However, the research could be strengthened by exploring the scalability of the proposed method, especially given the ever-increasing volume of data in digital forensics (Casey, 2011; Quick and Choo, 2014). At the same time, while the paper focuses on k-means and Spk algorithms, exploring the MAFS method's compatibility with other clustering algorithms could provide a more comprehensive understanding of its utility.

This research contributes significantly to the field of digital forensic analysis, particularly in efficiently clustering large volumes of crime-related documents. The proposed MAFS method's ability to improve clustering accuracy is of great importance for forensic investigators, aiding in quicker and more precise analysis of digital evidence. Future research could focus on the scalability of the MAFS method and its adaptability with other clustering algorithms, potentially broadening its applicability in various domains beyond digital forensics.

### 2.18. Fronesis: Pioneering Cyber-Attack Early Detection

The research paper titled *"Fronesis: Digital Forensics-Based Early Detection of Ongoing Cyber-Attacks,"* authored by Dimitriadis et al., (2023), presents a novel approach to detecting cyber-attacks in their early stages. presents a novel approach to detecting cyber-attacks in their early stages. The proposed ontology approach is displayed in *Figure 18*, and it is significant in the field of cybersecurity, as traditional detection methods relying on known signatures or machine learning often fall short against increasingly sophisticated cyberattacks. The authors rightly identify the gap in early detection of cyber-attacks, a crucial aspect considering that in 2020, only 59% of security incidents were detected by organisations themselves, with an adversary's median dwell time in a compromised system being 24 days (Mandiant, 2020). This context underscores the urgency of methods like Fronesis.

Its methodological strength lies in how it combines ontological reasoning with the MITRE ATT&CK framework and the Cyber Kill Chain (CKC) model, as well as using digital evidence from systems that are being watched. The application of rule-based reasoning to the Fronesis ontology for detecting cyber-attacks marks a significant advancement over traditional methods. By focusing on digital artefacts, which include both volatile data such as processes and non-volatile data like emails, Fronesis offers a more comprehensive detection mechanism. Section IV of the paper goes into more detail about how the Web Ontology Language (OWL) and the Semantic Web Rule Language (SWRL) are used to make the Fronesis ontology and its rule-based reasoning process work.

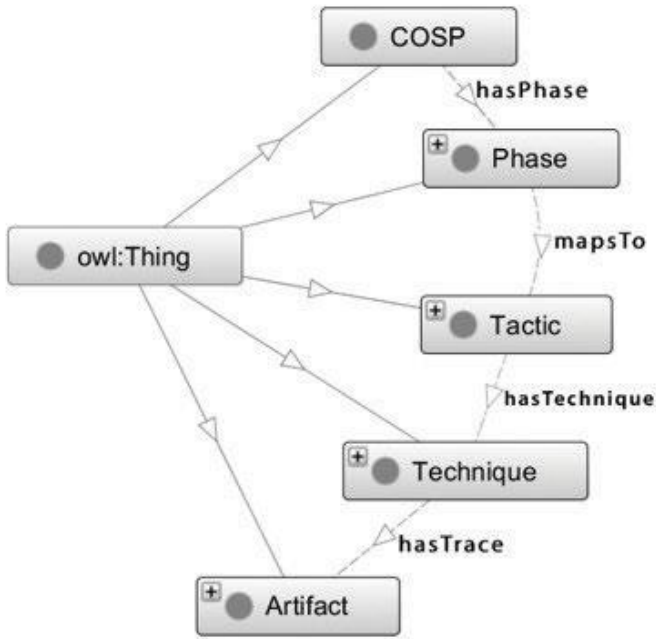

*Figure 18: OntoGraf rendering of the proposed ontology by Dimitriadis et al., (2023)*

The practical applicability of Fronesis is illustrated through an email phishing attack scenario. Phishing attacks, particularly prevalent in business environments, serve as a pertinent example to demonstrate the effectiveness of Fronesis in identifying and responding to real-world cyber threats. This example is instrumental in demonstrating the real-world applicability of Fronesis, emphasising its potential for improving cybersecurity defences.

The novelty of Fronesis lies in its multi-step methodology, combining the CKC model and MITRE ATT&CK, to reconstruct and detect a cyberattack based on digital artifacts. This approach surpasses the limitations of the CKC model by defining techniques for each CKC phase and using a wider array of digital artefacts for detection, leading to better results. Importantly, Fronesis focuses on detecting the cyberattack itself rather than just identifying specific techniques, thereby offering a more holistic and effective approach to cybersecurity.

While Fronesis represents a significant stride in cyber-attack detection, there are areas for further development. The reliance on digital artefacts, while comprehensive, also introduces the challenge of managing and analysing vast amounts of data. However, the paper could benefit from a broader evaluation of Fronesis across diverse attack scenarios beyond email phishing to demonstrate its versatility and robustness.

### 2.19. CQSS-GA-SVM: A New Era in Audio Forensics

The authors, Su et al., (2023), introduce a method for detecting and locating audio copy-move forgeries, utilising Constant Q Spectral Sketches (CQSS) combined with a custom genetic algorithm (GA) and support vector machine (SVM), as depicted in *Figure 19*. This method aims to address the challenges in blind audio forensics, particularly in identifying forgeries derived from the same audio recording.

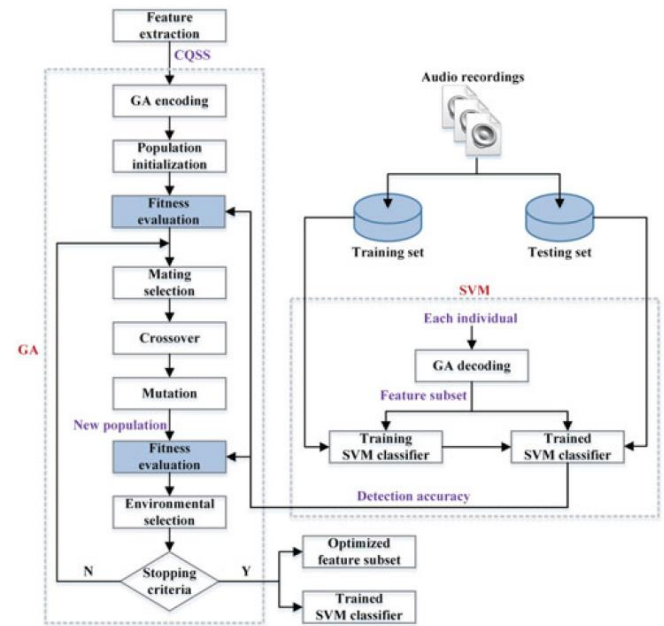

*Figure 19: Framework of the proposed CQSS-GA-SVM for the audio CMFD by Su et al., (2023)*

The integration of CQSS and GA-SVM represents a significant advancement in the field of audio forensics. The authors effectively extract CQSS features and then optimise these features using a custom GA combined with SVM. This approach not only enhances the detection accuracy but also automates the feature optimisation process, which is a notable contribution to the domain. The proposed method demonstrates high robustness against various post-processing-based anti-forensics attacks. This capability is crucial for practical applications in audio forensics, where tampered audio is often subjected to various manipulations to conceal the forgery.

The methodology shows adaptability to changes in duplicated segment duration, training set size, recording length, and forgery type. This flexibility is beneficial for forensic experts who deal with a wide range of audio forgery scenarios in real-world cases. The experiments conducted to evaluate the method's performance were thorough. The authors compared their approach against state-of-the-art methods, demonstrating its superiority in terms of accuracy and robustness. The use of real-world datasets for validation adds credibility to the results.

The statistical analysis provided in the paper offers a clear understanding of how CQSS features can capture subtle changes in forged recordings. This aspect of the research is well explained and contributes significantly to the validation of the proposed method. While the authors mention the method's efficiency, there is limited discussion on the computational complexity and actual processing time, which are critical factors in practical applications. Future work could focus on optimising the algorithm for faster processing without compromising accuracy. The paper primarily focuses on English and Chinese datasets. It would be beneficial to test the method's effectiveness across a broader range of languages and accents to ensure its applicability in diverse forensic scenarios.

As anti-forensics techniques continue to evolve, it is essential to regularly update and test the proposed method against newer forms of audio tampering. Continuous development in this area will maintain the relevance and effectiveness of the method.

### 2.20. Applying AI to Image Forensics

The paper titled *"The Application of Intelligent Systems to Digital Image Forensics"* by Lai and Chen (2009), which was presented at the Eighth International Conference on Machine Learning and Cybernetics, is a deep and thorough look at how genetic algorithms (GA) and support vector machines (SVM) can be used to find the camera that took a digital picture. This innovative research method, highlighted in the Research Flow Chart as seen in *Figure 20*, uses image features to determine the origin of digital images. The author has effectively employed genetic algorithms to automate the search for the most optimal features, along with the use of support vector machines for classifying these features. This approach is particularly significant in today's digital age, where the manipulation of digital images is a common challenge, thus calling into question the reliability of such images.

The results of the study underscore the effectiveness of the genetic algorithm in selecting fewer yet more pertinent features, achieving high rates of accuracy in identifying the source camera. This method marks a notable improvement over traditional techniques that often depend on metadata, which can be easily altered, or are less effective with high-end camera sensors. The findings are of considerable importance, offering a dependable method for digital image forensics, a crucial tool in various legal and security contexts.

The paper is commendable for addressing a relevant and timely issue in the field of digital forensics. The use of GA and SVM for image source identification introduces a novel approach to overcoming existing challenges in this domain. Moreover, the experimental design of the study, which includes the use of multiple cameras and a range of conditions like image resizing and the addition of post-processing graphics, adds a layer of robustness to the research. However, the study could benefit from an expansion of the experimental scope, incorporating a broader array of camera models and more diverse image manipulations.

The practical implications of this research are significant, especially in law enforcement and legal proceedings, where the authenticity of digital images is frequently a point of contention. The ability to trace an image back to its source camera accurately could revolutionise these fields. Nonetheless, the paper would benefit from discussing the limitations of this approach, particularly in terms of computational complexity and its efficacy against advanced image manipulation techniques. Future research could focus on integrating this method with other digital forensic techniques to enhance overall reliability.

### 2.21. Deep Learning Techniques

The publication *"A Survey on Deep Learning for Big Data"* by Zhang et al., (2018) covers many deep learning models and their use in big data. The authors divided the study into parts on common deep learning models like Stacked Auto-Encoders (SAE), Deep Belief Networks (DBN), Convolutional Neural Networks (CNN), and Recurrent Neural Networks.

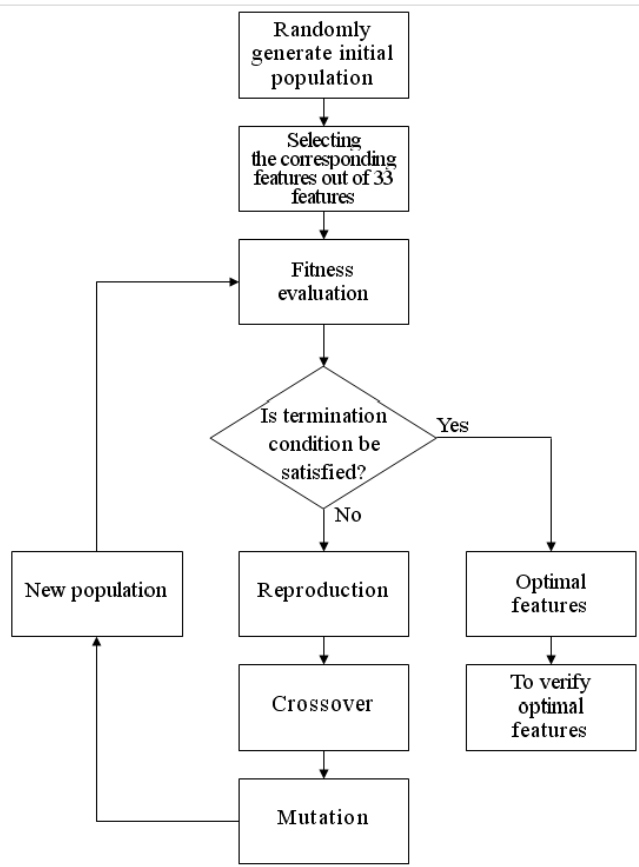

*Figure 20: Research Flow Chart by Lai and Chen (2009)*

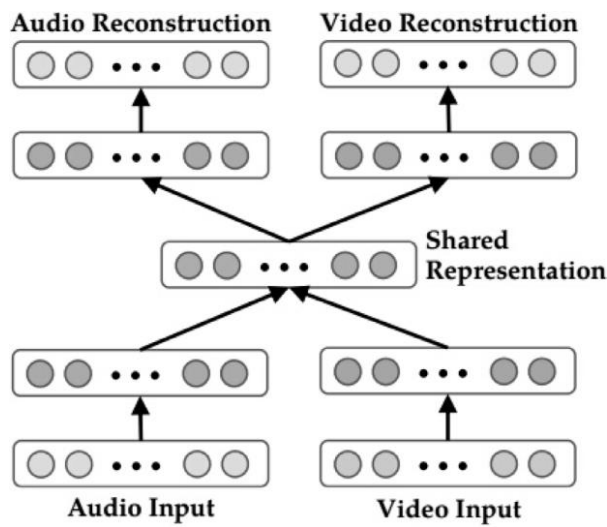

*Figure 21: Multi-Modal Deep Learning Model by Zhang et al., (2018)*

The architecture of the multi-modal deep learning model is depicted in *Figure 21.* The paper also discusses adapting these models to handle big data difficulties such as huge data volumes, heterogeneous data, real-time data processing, and low-quality data analysis. This structure shows how deep learning techniques for massive data management have evolved and adapted. A detailed exploration of deep learning architectures is the paper's strength. Each model is well explained, including its mechanics and applicability for certain big data applications. This degree of information helps readers comprehend how these models can be used in large-scale data situations.

However, the paper might use more case studies or real-world applications to demonstrate these models' practicality. While the theoretical and technical elements are thoroughly covered, practical examples would have helped explain how these models work in real life.

The work is well organised and walks the reader through deep learning with big data. The authors made difficult models understandable, yet some portions may require deep learning knowledge. The limitations and problems of these approaches when applied to big data seem underexplored. The research addresses issues like low-quality data, but a more in-depth analysis of these models' weaknesses would provide a more balanced picture.

**2.22. Deep Learning in Digital Forensic File Analysis**

The research paper *"Digital Forensic Analysis of Files Using Deep Learning"* by Mohammed et al., (2020) from Khalifa University offers an innovative approach to digital forensics. It focuses on the identification and analysis of file types within digital evidence, utilising deep learning to address the limitations of traditional forensic methods. *Figure 22* shows all the different file types used in the research.

| File Type | Dataset Size | Dataset Name |
|---|---|---|
| AVI | 6,766 | HMDB51 |
| MP4 | 20,338 | oops! |
| BMP | 3,000 | Caltech birds & NRC |
| JPG | 328,000 | MSCOCO |
| PNG | 1,000 | DIV2K |
| MP3 | 61,528 | Common Voice |
| WAV | 1,000 | RFS1k |
| PDF | 1,487 | GitHub responsivity |

*Figure 22: Different File Type used in the Research by Mohammed et al., (2020)*

The paper introduces a deep learning-based model for file type identification, marking a significant stride in addressing the challenges inherent in conventional forensic practices. Its innovative approach is particularly effective in accurately predicting corrupted files, a notable advancement over existing methods that often falter with such file types. This innovation and relevance are crucial in a field where accuracy and the ability to handle complex data are paramount.

In terms of methodology and technical rigor, the authors employed Convolutional Neural Networks (CNN), a choice backed by CNN's established efficacy in pattern recognition tasks. The structured methodology, encompassing dataset preparation, model training, and validation, contributes to the robustness of their approach. The inclusion of a diverse range of file types, such as AVI, MP4, JPG, BMP, and PNG, enhances the model's applicability across various digital formats.

The paper also presents a comprehensive validation of the model, demonstrating high accuracy rates in file type identification. These results underscore the effectiveness of the model. However, a more nuanced discussion regarding the model's limitations and the potential for false positives or negatives would offer a more rounded perspective on its real-world applicability. A thorough review of existing methodologies in digital forensics is also conducted, highlighting the inadequacies of traditional approaches and underscoring the necessity for a deep learning-based solution. This comparative analysis solidifies the argument for the proposed methodology and its potential to revolutionise digital forensic practices.

Lastly, the paper touches on the practical implications of this research for law enforcement and digital forensic experts. The prospect of deploying the model on portable devices for on-site examination opens up exciting possibilities for future work in the field. However, an exploration into the practical challenges of field implementation would have added valuable insights into the potential real-world impact of this research.

**2.23. Deep Learning in ImageNet Classification**

The paper *"ImageNet Classification with Deep Convolutional Neural Networks,"* researched by Krizhevsky et al., (2017), has made a profound impact in the domains of computer vision and deep learning. This critical review examines the paper's methodologies, innovations, results, and broader implications in the field. Krizhevsky et al., (2017) research introduces a deep convolutional neural network (CNN) that sets new benchmarks in processing large-scale image data. An illustration of the proposed CNN architecture by Krizhevsky et al., (2017) is depicted in *Figure 23*.

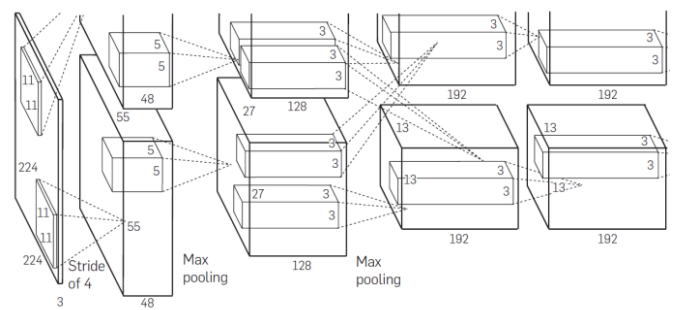

*Figure 23: An illustration of the proposed CNN architecture by Krizhevsky et al., (2017)*

Their network was trained on the ImageNet dataset and has five convolutional layers, followed by three fully connected layers. It was revolutionary at its inception and has since influenced a multitude of studies in deep learning. Rectified Linear Units (ReLUs) were a key innovation in their method because they made training deep networks faster than with traditional neuron models (Nair and Hinton, 2010). Also, using dropout as a regularisation method was a new idea at the time. It worked well to stop overfitting and showed how important it is for training big neural networks (Hinton et al., 2012).

The results of their work were nothing short of remarkable, with the network achieving a top-5 error rate of 15.3% in the ILSVRC-2012 competition, significantly outperforming the second-best entry. This achievement not only underscored the capabilities of deep learning in computer vision but also set a new standard for image classification tasks.

The paper's impact extends beyond its immediate results. It has demonstrated the feasibility of training deep neural networks on large-scale image datasets, inspiring a broad spectrum of research into CNNs and their applications across various domains (LeCun, Bengio and Hinton, 2015). This work established a foundational benchmark in image classification and catalysed further exploration in the field of deep learning.

However, the approach is not without its limitations. The substantial computational resources required for such deep learning models may limit their accessibility and practicality in certain contexts. The opaque nature of deep neural networks, as exemplified in Krizhevsky et al., (2017) research, poses challenges in terms of interpretability and understanding the decision-making processes within these models.

### 2.24. Impact of Physical Attacks on AI Systems

The research paper *"Robust Physical-World Attacks on Deep Learning Visual Classification"* by Eykholt et al., (2018), presented at the 2018 IEEE/CVF Conference on Computer Vision and Pattern Recognition, is a pioneering study in the field of AI security. A sample of physical adversarial examples against LISA-CNN and GTSRB-CNN is depicted in *Figure 24.*

*Figure 24: A sample of physical adversarial examples against LISA-CNN and GTSRB-CNN*

The paper explores the vulnerability of deep neural networks to adversarial examples in real-world settings, focusing on safety-critical applications like road sign classification in autonomous driving. It introduces the Robust Physical Perturbations (RP2) method, generating adversarial perturbations that can mislead classifiers under various physical conditions. The paper introduces the RP2 method, a pioneering approach in autonomous driving and other real-world DNN applications, addressing the gap in understanding adversarial examples in physical environments. This innovation is not only relevant but essential in today's rapidly evolving field of artificial intelligence.

The methodology employed in the paper further strengthens its impact. The authors use a two-stage evaluation, comprising both lab and field tests, to assess the effectiveness of adversarial attacks in real-world scenarios. This comprehensive and rigorous approach to testing underpins the credibility of their findings, demonstrating the robustness of RP2 in a variety of conditions. Such methodological rigour is commendable and adds significant value to the research. The practical implications of this research are profound, particularly for the safety and reliability of autonomous vehicles. The paper not only demonstrates the potential for adversarial attacks to cause misclassification of traffic signs but also highlights the urgent need for more resilient DNNs in safety-critical systems. This aspect of the research is particularly relevant in the context of the increasing integration of AI technologies in everyday life.

However, the research is not without its limitations. The scope of the study is somewhat narrow, focusing primarily on road sign classification. This limited focus presents an opportunity for future research to explore the application of RP2 in other domains where DNNs are used in physical environments, such as facial recognition systems, medical imaging technologies, and digital forensics investigations.

Moreover, the generalizability of the results could be a point of concern. The specific nature of the adversarial perturbations, such as stickers on stop signs, might limit the applicability of the findings to other scenarios. Future studies could investigate a broader range of physical-world attacks to enhance the universality and relevance of these insights in different contexts.

## 3. Comparison with Existing Literature Reviews

Goni et al., (2019) present a systematic review focusing on the application of machine learning (ML) algorithms in cybersecurity and cyberforensics. Their work highlights the critical aspects of confidentiality, integrity, and validity of data within cybersecurity. They delve into various cyber threats and the foundational concepts of ML algorithms, conducting a thorough literature review to explore their application in cybersecurity systems. In 2023, Balushi, Shaker, and Kumar emphasise the enhancement of digital forensic investigations through ML in their review paper. They discuss the various challenges encountered by forensic investigators and examine how ML algorithms aid in the analysis of digital evidence. Their paper categorises ML algorithms according to their specific uses in digital forensics and also discusses the limitations inherent in these technologies. Javed et al., in their 2022 survey paper, provide an extensive introduction to different computer forensic domains and tools. They conducted a comparative analysis of forensic toolkits and shed light on the current challenges and future research directions in computer forensics, adding a significant layer of understanding to this evolving field. Rizvi et al., (2022) explore the application of artificial intelligence (AI) in network forensics. Their research paper surveys existing methodologies, identifies the challenges faced, and suggests potential future directions in this niche but crucial area of cybersecurity. Kaur, Gabrijelčič, and Klobučar (2023) expand the discussion to the broader role of AI in cybersecurity. Their literature review analyses current research trends in AI and projects future pathways, offering a comprehensive view of AI's evolving influence in the cybersecurity landscape.

Our research contrasts with these existing literature reviews by offering a more holistic view that integrates insights from both AI and ML in the context of modern digital forensics and incident response. We provide a detailed analysis of the interplay and complementarity between AI and ML in digital forensics and incident response, an area not extensively explored in the other papers. Our paper addresses the practical applications of AI and ML in real-world incident response scenarios, shedding light on operational challenges and potential solutions. Furthermore, we present a forward-looking perspective, discussing emerging technologies and their potential impact on digital forensics and incident response.

While the existing literature reviews contribute valuable insights into specific aspects of cybersecurity, cyber forensics, and the roles of AI and ML, our research uniquely adds to the body of knowledge by offering a comprehensive, integrated, and practical perspective on the application of AI and ML in modern digital forensics and incident response. This holistic approach not only synthesises existing knowledge but also expands the discussion by highlighting practical applications and future possibilities in the field.

## 4. Comparative Analysis

This section provides a thorough comparison of the various AI and ML applications discussed in the literature review. Numerous investigation-improving applications are now feasible as a result of the incorporation of AI and ML techniques in digital forensics. The investigation of AI and ML applications has gained prominence as digital forensics professionals seek to remain ahead of evolving cyber threats. The comparative analysis reveals the transformative potential of AI and ML applications in digital forensics. Therefore, each application area presents unique contributions and difficulties, but collectively, they pave the way for more effective, precise, and proactive investigation techniques. By leveraging AI and ML, digital forensics professionals can not only keep pace with the ever-changing cyber landscape but also foster a culture of continuous development and innovation within the field.

| Category | Research Work | Contribution | Benefits | Drawbacks | Integration & Impact |
| --- | --- | --- | --- | --- | --- |
| Digital Forensic Investigation | (Song and Li* 2020) | Big data-driven forensic framework for high-volume, accurate, and secure investigations | Increases the likelihood of users receiving the desired results as per their commands | Does not address the issue of data validity and how it will be confirmed | Medium |
| Live Memory Forensics | (Thantilage and Le-Khac 2019) | Framework for extracting memory dumps from RAM for the purpose of acquiring forensic evidence. | Budget-friendly tool to help crime investigators self-analyse retrieved data. | Compatibility, data complexity, and performance issues | Medium |
| File Forensics | (Mittal and Memon 2021) | A new file identification method for data carving and memory forensics | Improved processing speed and functionality. | Overlooked the application of data-type classification | Low |
| Classification using Neural Networks | (Mohammad., 2018) | Method for using neural networks to evaluate, analyse, and draw conclusions from retrieved data for digital forensics. | Reconstruction of past events using neural networks. | The model may not be effective on certain types of data. | Medium |
| AI in Digital Forensics | (Hasan et al., 2011) | The model uses a database of crime scene evidence to train itself | Solve new and future crimes that may be strange to crime investigators | Vulnerable to hackers and third-party software | Low |
| Automated Artefact Analysis | (Du, Le and Scanlon, 2020) | Using Machine Learning to rank the order of priority of file artefacts needed for forensic examination | Reduce the psychological impact on human investigators who may be exposed to distressing evidence. | The system may be susceptible to over-fitting. | Medium |
| Network Analysis | (Ozcan et al., 2020) | Developed an edge-transmitted graph methodological approach for dealing with large forensic datasets | The framework can identify potential offenders by analysing the patterns of their email communication | The framework could be biased towards certain groups. | Medium |
| Data Mining & ML in Digital Forensics | (Tallón-Ballesteros and José, 2014) | Explored different data mining methods that investigators can apply to a digital forensic analysis and exammination. | The model can be used to assess large quantities of data faster and more accurately than human forensic. | Insufficient explanation of the model decision-making process. | Low |
| Machine Learning in Digital Forensics | (Toraskar et al., 2019) | Demonstrated SOM viability in criminal investigations. | SOM can be used to cluster data sets into groups that can be used to identify patterns of criminal activity. | Tools may be ineffective if a non-English language is used. | Medium |
| Live Memory Forensics Using ML | (Mosli et al., 2016) | Proposes a "heuristic approach" to automate malware detection | The approach can be used to detect malware already present in a system | Limitations in data visualisation and string processing | High |
| Live Memory Forensics | (Lashkari et al., 2020) | Uses a variety of classifiers to identify benign samples during binary classification. | Extracts essential features for malware analysis. | Risk of classifiers memorising too much noise. | Medium |
| Digital Forensic Investigation | (Casey, 2011) | Employs case studies to exemplify digital forensics utilisation in real-world investigations. | Valuable resource for individuals seeking comprehension of digital forensics. | Lacks coverage of advanced topics in digital forensics. | Low |
| Artificial Intelligence in Digital Forensics | (Du et al., 2020) | Highlight challenges and limitations in AI application to Digital Forensics, including requirements for substantial labelled datasets and potential biases. | AI enables the automation of manual tasks within digital forensics investigations, encompassing data analysis and categorization. | AI implementation in Digital Forensics investigations may engender ethical considerations. | Medium |
| Artificial Intelligence in Digital Forensics | (Qadir and Noor, 2021) | The framework relies on a deep learning algorithm trained on an extensive dataset of digital forensic artefacts. | The framework aids in the detection of overlooked evidence and novel patterns. | The framework's computational demands may limit its viability across diverse cases. | Medium |

*Table 3: Comparative Analysis of AI and ML Techniques in DFIR*

| Category | Research Work | Contribution | Benefits | Drawbacks | Integration & Impact |
| --- | --- | --- | --- | --- | --- |
| Philosophy and Machine Learning | (Thagard, 1990) | Lays crucial groundwork for future research into the philosophical and ethical ramifications of employing machine learning. | Automates various tasks, including data collection, analysis, and reporting, yielding time and resource savings for investigators. | Biases in machine learning algorithms may engender inaccurate outcomes. | Medium |
| Digital Forensics and Cybersecurity Statistics | (Insurance Information Institute, 2022) | Examine the ramifications of identity theft and cybercrime for individuals and organisations. | The report offers valuable insights into the scope and characteristics of identity theft and cybercrime. | The report's focus on the United States limits its generalizability to other nations. | Low |
| Data Science and Analysis | (Cabitza, Campagner, and Basile, 2023) | The authors distinguish between 'weak' and 'strong' perspectivist approaches. | Leverages uncertain and fuzzy data to enhance model performance, generalizability, and robustness. | The absence of a uniquely defined ground truth makes validation and evaluation more complex. | Low |
| Digital Forensics Research and Challenges | (Quick and Choo, 2014) | Identified the challenges posed by the increasing volume of digital forensic data and discusses the impact on the cost, complexity, and speed of digital forensic investigations. | A concise overview of the challenges posed by the increasing volume of digital forensic data. | Does not provide a detailed discussion of the technical or methodological aspects of digital forensic data analysis. | Medium |
| Digital Forensics and Cybercrime Legislation | (Mohammed et al., 2019) | Identify several future research challenges to address the rising volume of cybercrime in Nigeria. | Provides overview of the challenges faced by law enforcement officers investigating cybercrime. | Focusing on Nigeria jurisdiction limits its applicability to other countries. | Low |
| Forensic Trace Evidence Analysis | (Stoney and Stoney, 2015) | Propose a hybrid approach to forensic trace evidence analysis, integrating conventional and unconventional methodologies. | Concisely surveys the challenges in forensic trace evidence analysis. | Advocates a new approach to forensic trace evidence analysis but didn't elaborate on it. | Medium |
| AI Automation in Digital Forensics. | (Jarrett and Choo, 2021) | Present a concise review of the research on the impact of automation and AI on digital forensics. | Artificial intelligence and automation are poised to revolutionise digital forensics. | Ethical and legal considerations must be addressed before the full adoption of automation and AI in digital forensics. | Low |
| Big Data in Digital Forensics | (Guarino, 2013) | Concisely survey the challenges of digital forensics posed by the increasing volume of digital data. | Suggests some practical solutions to the challenges of digital forensics that could be beneficial to practitioners. | Limited availability and compatibility of forensic tools and techniques for big data analysis are potential limitations. | Medium |
| Online Predator Detection in Digital Forensics | (Ngejane et al., 2021) | Conducted a systematic review of machine learning-based approaches to online sexual predatory chat detection. | Offers a comprehensive overview of machine learning in detecting online sexual predatory chats. | Lacks a detailed discussion of the evaluation process used to assess the proposed machine learning algorithm. | Low |
| Introduction to Statistical Learning | (James et al., 2021) | Discussed the practical applications of statistical learning methods across domains. | Illustrates the material with numerous examples and exercises to facilitate readers' comprehension. | Does not specifically focus on digital forensics, readers may need to conduct further research to apply its concepts. | Low |
| Legal and Ethical Issues in Digital Forensics | (Losavio et al., 2018) | Exploration of unique legal issues arising from the deployment of IoT technologies in urban environments. | Provides potential conflicts and tensions between privacy, security, and DF investigations in smart cities. | Limited legal precedents and frameworks for addressing legal challenges related to IoT, DF, and security in smart cities. | Medium |
| Legal and Ethical Issues in Digital Forensics | (Brkan and Bonnet, 2020) | Discuss the legal and technical challenges of complying with the GDPR's requirement for explainable AI. | Valuable resource for those interested in the legal implications of algorithmic decision-making in DF. | Exploration of ethical concerns and biases in interpreting and communicating algorithmic decision explanations. | Medium |
| Deep Learning and Big Data Analytics | (Zhang et al., 2018) | Deep learning algorithms and methodologies for handling big data. | State-of-the-art deep learning research for big data, which can help readers stay up-to-date on recent advances. | Does not discuss recent deep learning developments for big data, such as GANs and NLP. | Low |

| Category | Research Work | Contribution | Benefits | Drawbacks | Integration & Impact |
| --- | --- | --- | --- | --- | --- |
| Big Data Analytics in Digital Forensics | (Mohammed et al., 2016) | Evaluated on a real-world dataset, with results showing effectiveness in identifying and extracting digital evidence from heterogeneous big data. | Reviews the challenges of digital forensics analysis of heterogeneous big data and how the authors' approach addresses them. | Reviews the challenges of digital forensics analysis of heterogeneous big data and how the authors' approach addresses them. | Low |
| Digital Forensics and Image Analysis | (Krizhevsky et al., 2017) | The CNN achieved an 84.6% classification accuracy on the ImageNet database, a significant improvement over previous methods. | Advancements in deep learning techniques and the use of large-scale datasets, such as ImageNet, for training deep neural networks. | CNNs are computationally demanding to train and deploy. | Medium |
| AI in Digital Forensics and Incident Response | (Dunsin et al., 2022) | Present an exploration of how AI techniques can be applied in the field of DF, in the context of incident response in a constrained environment. | Present how AI techniques can be tailored and optimised to work effectively in resource-constrained environments. | Ethical and privacy concerns related to the use of AI in digital forensics and incident response | Medium |
| Adversarial ML in Digital Forensics | (Biggio et al., 2013) | Provides a comprehensive overview of evasion attacks on machine learning models at test time. | Helps researchers understand the challenges of deploying ML models in real-world scenarios. | Lacks empirical evidence to support the effectiveness of the proposed methods. | Medium |
| Model Interpretability | (Lipton, 2018) | Exploration of the interpretability limitations and trade-offs in complex models | Highlighting the complexities and limitations of model interpretability | Interpretability of complex models, such as deep neural networks with millions of parameters, is challenging. | Low |
| Live Memory Forensics | (Kiley, Dankner, and Rogers 2008) | Propose techniques and methods for recovering volatile data from instant messaging applications. | Enhance investigators capabilities to retrieve and analyse evidence from instant messaging applications. | Did not address the evolving nature of instant messaging applications and their security and forensic challenges. | Low |
| File Carving | (Sari and Mohamad, 2020) | File carving techniques and review graph theoretic and weightage techniques. | Provides a taxonomy of graph theoretic and weightage techniques for comparing and contrasting different techniques. | Neglects to address the ethical implications of using graph theory and weightage techniques in file carving. | Medium |
| File Fragment Classification | (Teimouri et al., 2020) | Fragments-Expert toolbox with a graphical user interface for classifying file fragments. | Fragments-Expert's graphical user interface simplifies file fragment analysis for non-experts and less experienced users. | The effectiveness of Fragments-Expert in classifying file fragments may be affected by the complexity of the data and the diversity of file types. | Medium |
| Digital Forensics Timeline Analysis and Knowledge Representation | (Chabot et al., 2014) | formalised knowledge representation model for advanced digital forensic timeline analysis. | The model can identify relationships and dependencies among digital artefacts, which facilitates understanding the chronological order and context of events. | Proposed model's practical implementation and scalability in real-world digital forensic investigations may need further exploration and validation. | Low |
| Machine Learning in Digital Forensics | (Nila et al., 2020) | Propose a machine learning-based quick incident response approach. | The approach can rapidly and accurately classify new incidents. | The approach relies on a large dataset of known incidents to train the model. | Medium |
| Adversarial Attacks and Defenses | (Eykholt et al., 2018) | Evaluated the effectiveness of proposed attacks on deep learning models and identified their vulnerabilities. | Identifies limitations of current deep learning algorithms and advances the field by suggesting areas for further research and development. | The evaluation results may vary depending on the deep learning models and datasets used or real-world scenarios. | High |
| Data Analysis and Bias Mitigation | (Balayn et al., 2021) | Identifying and mitigating bias and unfairness in the digital lifecycle stages | Provides guidelines and best practices for ensuring fairness and mitigating biases in data-driven decision support systems. | Proposed approaches to specific Digital Forensics scenarios may require further exploration and validation. | Medium |
| Chain of Custody Management in Digital Forensics | (Elgohary et al., 2022) | Propose a method for improving the reliability of the chain of custody for image forensics investigation applications. | The method can improve the accuracy of image forensics investigations. | The method requires a large image dataset for training the mathematical model. | Low |

## 5. Discussions

On the basis of the findings of the systematic literature review, it is strongly recommended that the digital forensics community continue to embrace and research the benefits of artificial intelligence (AI) and machine learning (ML). Significant investments in the development and implementation of sophisticated tools and applications are required to fully leverage these technologies. These resources are necessary to effectively support the growth and expansion of digital forensic technologies, processes, and procedures. Digital forensics experts should investigate the use of artificial intelligence techniques such as pattern recognition, expert systems, and knowledge representation. These techniques have the potential to significantly enhance the efficiency and effectiveness of cybercrime investigation capabilities and processes.

It is necessary to resolve a number of obstacles associated with the adoption of AI techniques in order to ensure their optimal application. Due to the vast quantities and intricacy of data generated by online activities, scalability is a crucial concern. It is possible to make AI techniques much more useful in digital forensics by making them better at handling and processing such huge amounts of data. Equally, the admissibility of AI-collected evidence in court proceedings is contingent on its reliability being established. This can be achieved by instituting standardised procedures and guidelines that facilitate the admissibility of digital evidence in legal settings.

There is a need for numerous, distinct, and extensive studies that can help address the ongoing issues and deficits in forensic examinations. These studies should concentrate on the development of more efficient and effective AI techniques capable of addressing the unique challenges faced by professionals in digital forensics. During the analysis and examination phases of the digital forensics lifecycle, it is also crucial to investigate novel AI and ML application domains. Notably, digital forensics must address two major issues: malware infection investigation and Windows registry forensics. Developing comprehensive AI-based strategies to resolve these issues will enable the detection and analysis of malware infections as well as the extraction and interpretation of pertinent information from Windows registry files.

Furthermore, it is crucial to maintain vigilance when monitoring and assessing the use of AI and ML techniques in digital forensics. This governance is required to ensure that their deployment is ethical, transparent, and respectful of data subjects' rights and privacy. Constant evaluation and assessment aid in identifying potential risks or negative outcomes associated with the use of these technologies, thereby enabling prompt mitigation measures.

The systematic literature review concludes by highlighting the importance of AI and ML in digital forensics and recommending their continued application. For AI techniques to attain their full potential in this field, it is necessary to address scalability and evidence validation concerns, conduct exhaustive research, and monitor their ethical application.

Also, when it comes to computer forensics, particularly Windows-based investigations, it is imperative to address not only Windows Registry forensics but also the examination of a variety of other artefacts such as System Resource Usage Monitor (SRUM), Prefetch, AmCache, and others. While the Windows Registry is a critical component containing vital system and application settings, these other artefacts offer a wealth of information about system usage, application execution, and file activities. AI and ML techniques can be particularly advantageous in parsing, analysing, and correlating data from these diverse sources. By extending AI's application beyond Windows Registry forensics to include these artefacts, digital forensic professionals can achieve a more comprehensive understanding of system interactions and user activities, leading to more robust investigations. This holistic approach acknowledges the complex nature of Windows forensics and leverages the full spectrum of data available in Windows environments for more effective and efficient investigations. Nonetheless, robust research and collaboration within the digital forensics community are required to advance the field and overcome these obstacles.

## 6. Conclusions and Future Research Directions

Based on the exhaustive survey conducted in this study, a strong recommendation emerges for the integration of artificial intelligence (AI) and machine learning (ML) methodologies in ongoing and future digital forensics research. These techniques hold significant potential for enhancing investigative precision and efficacy, particularly in addressing the escalating prevalence of cybercrime. Nonetheless, the issue of data validity demands careful attention, especially when dealing with diverse data from various individuals, devices, platforms, and cultural contexts. Achieving higher success rates necessitates the formulation of precise objectives, utilising AI systems and data sources tailored to specific regions.

Recent insights from memory forensics underscore the need for cautious consideration when selecting tools for memory dump retrieval, weighing their strengths and limitations. Expanding the scope of research to encompass mobile devices and intelligent home appliances represents a logical progression to address the evolving landscape of digital evidence sources.

By making AI and ML applications more powerful, refining pre-processing techniques and email address matching, and using data from many different sources, potential offenders can be found much more accurately. The incorporation of various analysis experiments for comparison purposes promises improved data mining techniques and reduced errors during the analytical process. Embracing machine learning-based metadata analysis, the utilisation of multiple cluster sizes, and the adoption of self-organising maps can enhance the precision of results and contribute to the development of innovative methodologies.

A pivotal gap exists in malware artefact detection within current forensic investigations. In response, this study proposes the application of Reinforcement Learning, modelling it as a Markov decision process (MDP), aligning with RL's strength in capturing intricate agent-environment dynamics. Integrating this into a comprehensive framework offers a path towards automated malware detection. The transition matrix diagram serves as a visual representation, aiding comprehension of potential transitions and their probabilities within the MDP and guiding the construction of robust malware detection models.

Finally, this study underscores the vitality and practical applicability of AI in the realm of digital forensics. The adoption of AI techniques promises swifter and more effective investigations, facilitated by the identification of data patterns indicative of cybercrime and potential culprits. While AI techniques such as pattern recognition, expert systems, and knowledge representation contribute significantly to combating cyber threats, the evolving nature of data representation necessitates adaptable methods. Recognising the limitations of existing approaches and addressing scalability concerns within a legal framework is paramount. This demands a concerted focus on the development of tools and applications that harness the full potential of AI in digital forensics while ensuring ethical and legally admissible processes.